# *Rendezvous Regions*: A Scalable Architecture for Resource Discovery and Service Location in Large-Scale Mobile Networks

Karim Seada, Ahmed Helmy
Electrical Engineering Department, University of Southern California
{seada, helmy}@usc.edu

***Abstract:*** *In large-scale wireless networks such as mobile ad hoc and sensor networks, efficient and robust service discovery and data-access mechanisms are both essential and challenging. Rendezvous-based mechanisms provide a valuable solution for provisioning a wide range of services. In this paper, we describe Rendezvous Regions (RRs) - a novel scalable rendezvous-based architecture for wireless networks. RR is a general architecture proposed for service location and bootstrapping in ad hoc networks, in addition to data-centric storage, configuration, and task assignment in sensor networks. In RR the network topology is divided into geographical regions, where each region is responsible for a set of keys representing the services or data of interest. Each key is mapped to a region based on a hash-table-like mapping scheme. A few elected nodes inside each region are responsible for maintaining the mapped information. The service or data provider stores the information in the corresponding region and the seekers retrieve it from there. We run extensive detailed simulations, and high-level simulations and analysis, to investigate the design space, and study the architecture in various environments including node mobility and failures. We evaluate it against other approaches to identify its merits and limitations. The results show high success rate and low overhead even with dynamics. RR scales to large number of nodes and is highly robust and efficient to node failures. It is also robust to node mobility and location inaccuracy with a significant advantage over point-based rendezvous mechanisms.*

## 1 Introduction

Current research in infrastructure-less wireless networks can be categorized into two main categories: mobile ad hoc networks and sensor networks. There are many similarities between the two categories, but the major challenges are typically different. For efficient service provisioning, the challenges in ad hoc networks are the lack of infrastructure and the highly dynamic nature of nodes and their unpredicted mobility patterns. While in sensor networks the challenges are mainly the limited resources and the extremely large number of nodes. Some applications of sensor networks involve also mobility. In sensor networks, each device is capable of some computation, wireless communication, and sensing under energy-constrained conditions. Communication in sensor networks is typically application-specific and data-centric, and it consists of the tasks sent to nodes and the data recorded by nodes about the environment. Typical approaches for locating resources and data items in these networks rely on either flooding or centralized external storage. Both could suffer from scalability and efficiency problems. In this paper, we describe Rendezvous Regions (RRs) - a novel self-configuring, scalable, efficient and robust rendezvous-based architecture.

In our architecture, the network topology space is divided into rectangular geographical regions, where each region is responsible for a set of keys representing the data or resources of interest. A key, $k_i$, is mapped to a region, $RR_j$, by using a hash-table-like mapping function, $h(k_i)=RR_j$. The mapping is known by all nodes and is used during the insertion and lookup operations. A node wishing to insert or lookup a key obtains the region responsible for that key through the mapping, then uses geographic-aided routing to send a message to the region. Inside a region, a simple local election mechanism dynamically promotes nodes to be servers responsible for maintaining the mapped information. Replication between servers in the region reduces the effects of failures and mobility. By using regions instead of points, our scheme requires only approximate location information and accordingly is more robust to errors and imprecision in location measurement and estimation than schemes depending on exact location information. Regions also provide a dampening factor in reducing the effects of mobility, since server re-election is not invoked as long as current servers move inside their regions and hence the overhead due to mobility updates is quite manageable.

We run extensive detailed simulations to investigate the design space, and study the architecture in various environments including node mobility and failures. In addition, we perform high-level simulations and analysis to analyze RR scalability and evaluate it against other approaches; flooding, centralized storage and GHT [17], to identify its merits and limitations. The results show that RR is scalable to large number of nodes and is highly efficient and robust with node mobility, failures, and location inaccuracy. We like to emphasize that our goal from the comparison is not to say that one approach is always better, but to show the strengths and limitations of different approaches and under which conditions and environments each is preferable.

The rest of the paper is outlined as follows. In Section 2 we discuss related work. In Section 3 we provide the context and assumptions under which our architecture operates. Section 4 explains the design and section 5 contains the detailed evaluation of the architecture. Conclusions are presented in Section 6.

## 2 Related Work

In wireless networks, the simplest form of data dissemination or resource discovery is global flooding. Flooding does not scale well. Other approaches that address scalability employ hierarchical schemes based on cluster-heads or landmarks [12]. These architectures, however, require complex coordination between nodes, and are susceptible to major re-configuration (e.g., adoption, re-election schemes) due to mobility or failure of the cluster-head or landmark, incurring significant overhead. GLS [13] provides a scalable location service by using a predefined geographic hierarchy and a predefined ordering of node identifiers to map nodes to their locations. GLS is presented for locating nodes and assumes that node identifiers are known.

In sensor networks communication is identified as data-centric based on the content of data rather than node identities. A data-centric routing scheme presented is directed diffusion [8]. Directed diffusion uses flooding to advertise the interests from sinks to sources throughout the network. We believe that our rendezvous-based architecture can be integrated with directed diffusion to discover resources in a scalable manner instead of using flooding mechanisms.

The original RR idea borrowed from our earlier work on PIM-SM rendezvous mechanism [4] that uses consistent mapping to locate the rendezvous point (RP). However, a rendezvous *point* is insufficient in a highly dynamic environment as wireless networks. We first hinted at the RR idea in [6], in the context of bootstrapping multicast routing in large-scale ad hoc networks, with no protocol details or evaluations. In this work, we present a detailed architecture for RR, with full description of the design and the mechanisms to deal with mobility, failures, and inaccuracies, and generalizing it to deal with resource discovery and data-centric architectures in general.

Our architecture requires nodes to know their approximate locations. Location-awareness is essential for many wireless network applications, so it is expected that wireless nodes will be equipped with localization techniques. In general, many localization systems have been proposed in the literature: GPS, infrastructure-based localization systems [22][16], and ad-hoc localization systems [3][18]. For an extensive survey of localization refer to Hightower *et al.* [7]. In all these localization systems an estimation error is incurred that depends on the system and the environment in which it is used. GPS is relatively accurate, but it requires visibility to its satellites and so is ineffective indoors or under coverage. In addition, the high cost, size, and power requirements make it impractical to deploy on all nodes. Infrastructure-based localization systems are mostly designed to work inside buildings and they either have a coarse-granularity of several meters or require a costly infrastructure. Ad hoc localization systems can have high localization errors due to environmental factors affecting the location measurements. In our design we attempt to provide an architecture that requires only approximate location information.

Several geographic routing protocols (e.g. [9][11]) have been proposed. GPSR [9] is a geographic routing protocol for wireless networks that works in two modes: greedy mode and perimeter mode. In greedy mode, each node moves the packet closer to the destination at each hop by forwarding to the neighbor closest to the destination. Greedy forwarding fails when reaching a dead-end (local maximum), a node that has no neighbor closer to the destination. Perimeter routing (face routing) is used to route around dead-ends until closer nodes to the destination are found. In perimeter mode, a packet is forwarded using the right-hand rule in a planar embedding of the network graph. Since wireless network connectivity is in general non-planar, each node runs a local planarization algorithm such as GG or RNG, to discard a subset of the physical links during perimeter routing, so that the resulting graph is planar.

As mentioned earlier, we presented a high-level description of the original RR idea in [6]. Later on, in GHT [17] a related idea was proposed for data-centric storage in sensor networks. GHT is a geographic hash table system that hashes keys into geographic *points*, and stores the key-value pair at the sensor node closest to the hash of the key. GHT requires nodes to know their exact geographic location and uses the GPSR [9] protocol, to reach the destination. GHT uses GPSR perimeter routing to identify a packet home node (the node closest to the geographic destination). Packets enter perimeter mode at the home node (since no neighbor could be closer to destination), and traverse the perimeter that enclose the destination (home perimeter) before returning back to home node. GHT uses a perimeter refresh protocol to replicate keys at nodes in the home perimeter. The perimeter refresh protocol uses perimeter routing to refresh keys periodically, in order to detect topology changes after failures or mobility.

The obvious distinction between RR and GHT is using a rendezvous region instead of a rendezvous point. However, the design goals and architectural details are quite different. GHT was designed for sensor networks with low mobility, while a main goal in RR design is to target also high mobility environments. RR is also based on our objective to design geographic systems that need only approximate location information. The use of regions affects many design details such as the server election, insertion, lookup, and replication, as will be explained in the design section.

## 3 Context

In this section, we present the data models we consider in our study. We introduce two models, the *service model* and the *event model.* We then give a brief statement about the geographic requirements of Rendezvous Regions and present our assumptions.

### 3.1 Data Model

RR provides a general architecture for data-centric storage and resource discovery. Data operations could be viewed as general insertions and lookups of keys. Different applications differ in the number and characteristics of their insertions and lookups. The ratio between lookups and insertions affects the performance of our architecture; we call it *LIR* (Lookup-to-Insertion Ratio). We will define two models and identify data services and applications suitable under these models.

#### *3.1.1 Service Model*

In this model, the number of lookups is much larger than the number of insertions. Hence, this model has large LIR. Stored data are long-lived and queried continuously by large number of nodes. Any node in the network can perform insertion or lookup. This model is viewed as a general service location model and several applications could be mapped to it:

- Service location & bootstrapping: In ad hoc networks, nodes can use RR to identify and find new services. The service provider maps its service to a region ($RR_i$ = *hash(service)*) and stores information about the service (e.g. location) there. Nodes interested in the service use the same hash function to get the information from that region.
- Users and object tracking: Users (or tracked objects) in the ad hoc network map their identities to the corresponding region and store their location there. Other users use the same mapping to retrieve the location from that region.
- Configuration & task assignment in sensor networks: During the startup of the network operation, it is assumed that nodes' configuration and tasks are flooded to the entire network. Global flooding is inefficient if nodes are not joining simultaneously, new nodes are added, or tasks of different nodes are changing with time. RR provides a flexible way for nodes to get this information. Each node will get its task from a rendezvous region computed as a hash of the node location or other properties of the node environment (e.g. $RR_i$ = *hash(location or properties)*). Operators assign tasks to nodes by sending them to $RR_i$ using the same hash function.
- Data-centric storage for events that do not change frequently, such that the number of detected events is lower than the number of queries sent by nodes. For example, detecting a certain phenomenon in an area, which is required by other sensors to perform their computations, or required by actuating devices in order to decide on certain actions.
- Providing global properties of the network or environment that are of interest to many nodes.
- Database querying: Several research is going on viewing the sensor network as a database and using declarative SQL-like queries for retrieving information [5][1]. RR can provide an in-network storage component in these schemes. For example, temperature sensors send the temperatures of their regions to '$RR_i$ = *hash(temperature)*' and a node interested in the max temperature sends a query 'Select MAX temperature' that retrieves this information from $RR_i$.

#### *3.1.2 Event Model*

This model has large number of insertions with a lower number of lookups. An example is data-centric storage with many events detected from a set of event types. A query looks for the events detected of a certain type. This is similar to the model presented in GHT [17]. Each event type is hashed into a rendezvous region and all events detected of that type are forwarded there. Events detected cause insertions, while queries for events are the lookups. Aggregation could be valuable in this model, where a query for a certain type returns an aggregation of all events detected for that type. We assume that any node in the network detects and inserts events, while queries are sent by external nodes through specific access points.

### 3.2 Geographic Requirements

A main objective of our architecture is to *relax the requirements for exact geographic information*. Nodes need only to know their regions and so the exact location may not be required. We design our system as an *overlay* that can run over different routing protocols. The routing protocol need not be a geographic routing protocol that requires the exact geographic positions of nodes. It can be any routing protocol augmented to provide approximate routes toward regions. Relaxation of the dependence on geographic information is valuable for many reasons. Geographic information may not be always available and it is not accurate. Obstacles and non-ideal radio ranges can cause many problems to current geographic routing schemes. In addition, mobility increases the inaccuracy of location updates.

In our recent studies, we show how location errors, caused by localization systems and inaccuracy [20],

inconsistency of location dissemination [10], or node mobility [21], result in severe performance degradation and correctness problems in geographic routing protocols as GPSR [9] and GOAFR [11], in addition to GHT [17]. It is therefore crucial to provide alternative architectures that relax the assumptions of availability of accurate location information. Our Rendezvous Regions approach provides one such architecture and is more robust to errors and imprecision in location measurement and estimation than schemes depending on exact geographic location. In the results section, we show the effect of location inaccuracy on RR and GHT.

### 3.3 Assumptions

The network space is divided into rectangular equal-sized regions (see Figure 1), where the size of the region is set based on the radio range and how many hops we want the region to cover. The region size we use is covering a few radio hops, to provide an adequate relaxation of the inaccuracy and mobility effects, while keeping the region flooding overhead and server load reasonable. In the simulations we study the effect of the region size in more detail. We assume that the geographic space and boundaries of the network are known and that each node has a localization mechanism to detect its approximate geographic location and accordingly its region. Our design also allows us to relax the requirements for exact boundaries by having boundary regions instead of boundary points. Since nodes know the network geographic space boundaries and the region size, they can determine their regions within the space. Using appropriate density, we assume that the network is connected and that each region has nodes in it, which is a valid assumption under a reasonable density. In case of partitions and empty regions, multiple hash functions can provide substitute regions, when the original region is not available.

## 4 Design Overview

The network topology space is divided into geographical regions (*RRs*), where each region (e.g., $RR_j$) is responsible for a set of resources. The resource key space is divided among these regions, such that each resource key ($K_i$) is mapped to a region. The key-set to *RR* mapping ($KSet_i \leftrightarrow RR_j$) is known by all nodes. Inside each region, a set of elected servers is responsible for holding these resources' information. We introduce additional mechanisms to maintain consistency in case of mobility or failures.

The Rendezvous Regions scheme can be built on top of any routing protocol that can route packets toward geographic regions. The only requirement of the routing protocol is to maintain approximate geographic information, such that given an insertion or lookup to a certain region, it should be able to obtain enough information to route the packet toward that region. Given that the packet is able to reach the region, there are several design options inside the region itself. These design options affect the operation of insertions, lookups, server election, and replication inside the regions. They also affect the consistency operations for mobility and failures. In our experience, it is usually the case that no one design option gives best performance under all operating conditions. The design choices depend on many factors such as the environment, application, and the pattern of insertions and lookups. In this work we attempt to investigate and study several design options to identify their strengths and weaknesses. There are mainly three options for forwarding packets inside the region: (a) Geocast, (b) Anycast, and (c) Unicast.

(a) *Geocast*: By geocast we mean sending the packet to all nodes in a geographical area. Geocasts suffer from a high overhead, but are practical when we want to send the packet to several non-determined nodes (in our case to the servers) within a specific geographic region. It is also robust in the face of dynamics and does not depend on the underlying routing protocol

(b) *Anycast*: Anycasts are used when it is sufficient to reach any node of a set of nodes (any server). It has similar advantages to geocasts and can be implemented by using expanding ring search techniques or by caching previously known servers.

(c) *Unicast* to servers: Direct unicasts can be used when the locations of servers are well-known, either because the servers are fixed or because they coordinate to keep their information up-to-date. This has the advantage of low overhead during insertions and lookups, but if the servers are dynamic, extra periodic overhead for maintaining the servers' information is required. Unicasting may also depend on the underlying routing protocol to keep track of nodes' locations. The additional periodic overhead can be acceptable if the insertion or lookup rates are high.

In our current design we use geocasts for insertions and anycasts for lookups. These design choices are simple to implement, robust to dynamics, and do not require tracking of nodes' locations. We plan to investigate other server coordination techniques in future work. Following we describe the main components of our architecture.

### 4.1 Region Detection

Using a localization mechanism [7], each node detects its location and accordingly its geographic region. When the node moves, it detects its new location and so it can keep track of its region. The node uses this information to forward packets toward regions, to detect packets forwarded to its region, and to potentially participate in server election in its region (if and when needed) (Figure 1).

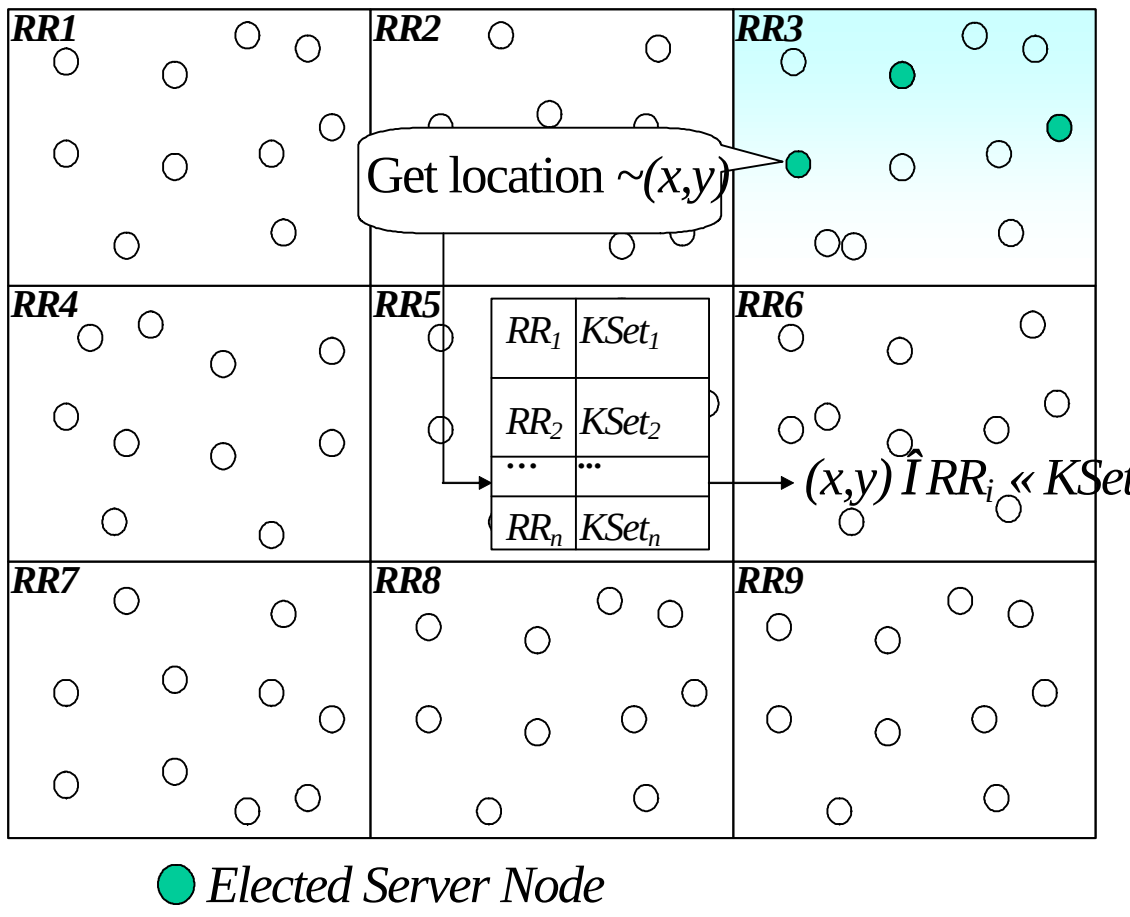

Figure 1: By knowing their approximate location ~($x$,$y$), and hence region ($RR$), nodes in every region (shown in $RR3$) conduct local elections. Only a few nodes (the elected servers) maintain information about the resource keys ($KSet_i$) mapping to $RR3$.

## 4.2 Server Election

A simple local election mechanism is used inside the region to dynamically promote the servers. The number of servers required in the region is called $S$. As $S$ increases, the robustness to mobility and failures increases, but also the storage overhead increases. In the simulations we study the effect of $S$ in more detail. Servers are elected on-demand during insertions. When a data insertion operation is issued, the first node in the region that receives the insertion[1], known as the *flooder*, geocasts the insertion inside the region. Each server receiving the insertion geocast sends an Ack back to the flooder. The flooder keeps track of the servers and if it does not get enough Acks (the minimum number of servers required), it geocasts again and includes a self-election probability, $p$, in the goecast message. Each node receiving the geocast elects itself with probability $p$ and if it becomes a server, it replies to the flooder. If not enough Acks are received, the flooder increases $p$ based on a back-off mechanism until the required number of servers reply or $p$ reaches 1. When servers move out of the region or fail, new servers are elected in the same way. After the new servers are elected, they retrieve the stored keys from other servers.

## 4.3 Insertion

A node inserts a key, $K$, by first mapping the key to a rendezvous region, $RR_i$, where $K \in KSet_i \leftrightarrow RR_i$. The node generates a packet containing the region identifier, $RR_i$, in its header. Nodes routing the packet toward the region, check the region identifier to determine whether they are in or out of region. The first node inside $RR_i$ to receive the packet, the *flooder*, geocasts the packet inside the region. Servers inside the region receive the geocast, store the key and data, then send Acks back to the flooder (Figure 2). The flooder collects the Acks and sends an Ack back to original sender. If no Ack is received by the sender, it timeouts and retransmits the insertion up to a fixed number of times.

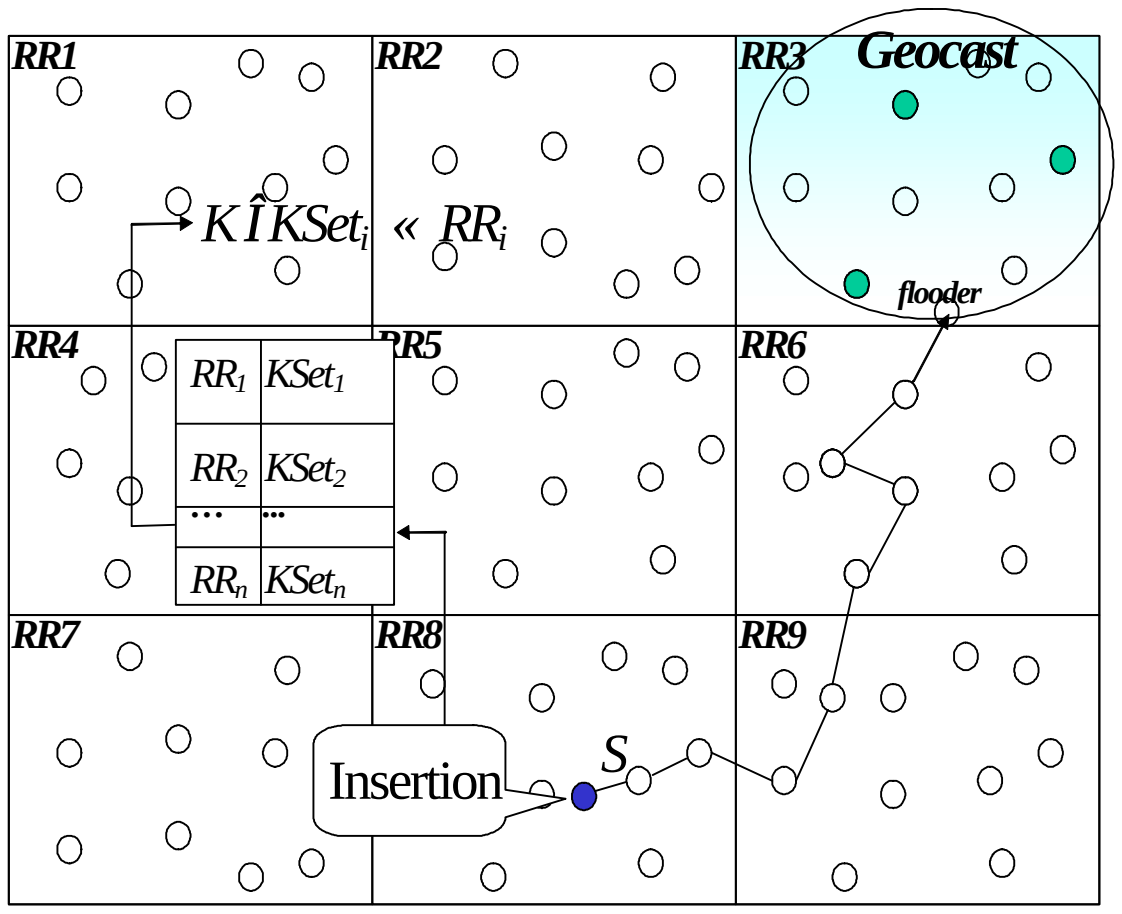

Figure 2: Insertion (step I): Node $S$ wishing to insert (or store) resource key $K$ that belongs to $KSet_i$ gets the corresponding $RR$ (in this case $RR3$) through the mapping ($KSet_i \rightarrow RR_i$). (step II): Node $S$ sends the resource information towards $RR3$, where it is geocast by the flooder and stored by the servers.

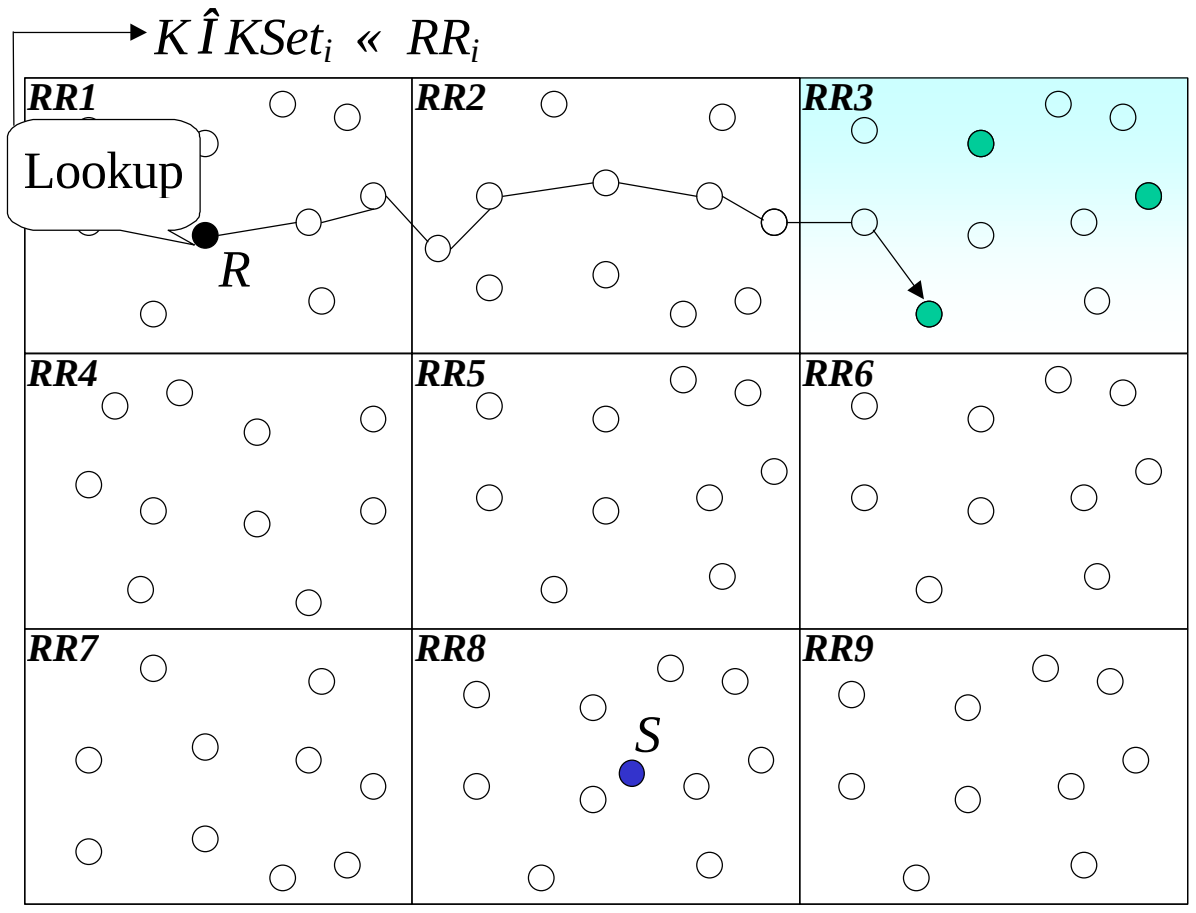

Figure 3: Lookup: Node $R$ looking for a resource with key $K$ that belongs to $KSet_i$ gets the corresponding $RR$ (in this case $RR3$) through the mapping ($KSet_i \rightarrow RR_i$). Then sends the resource lookup towards $RR3$, where it is anycast to any server holding the information.

## 4.4 Lookup

Lookups are similar to insertions except that nodes and previous flooders inside a region cache locations of the recent servers they hear from, and send the lookups directly to any of the servers (anycast). The server replies to the flooder and the flooder replies back to original sender (Figure 3). If the flooder receives no reply or if it has no cached servers, it geocasts the lookup inside the region.

[1] A node can identify that it is the first node in the region to receive the packet by a simple flag set in the packet header.

### 4.5 Replication

Replication is inherent in this architecture, since several servers inside the region store the key and data. This adds extra robustness to failures and mobility. For additional robustness against severe dynamics such as group failures and partitions, multiple hash functions may be used to hash a key to multiple regions.

### 4.6 Mobility

Local movements of nodes and servers have negligible effect and overhead on our architecture as long as servers stay within their regions. The only condition we need to consider is when a server moves out of its region. The server checks its location periodically to detect when it gets out of its region, in order to send an insertion packet toward that region so that new servers are elected. The server then deletes its stored keys and is not a server anymore. It may or may not get elected again later in a new region.

### 4.7 Failures

Since each region contains several servers, and insertions and mobility may invoke new server elections, it is unlikely that independent reasonable failures will cause all servers to vanish. In order to avoid this case anyway, servers use a low-frequency periodic soft-state mechanism during silent (low traffic) periods, to detect failing servers and promote new servers. Each server runs a low-frequency timer, which is reset each time an insertion geocast is received. When the server times out, it geocasts a packet checking for other servers. Other servers reset their timers upon receiving this check and reply back demonstrating their existence. If not enough servers reply back, server election is triggered.

### 4.8 Bootstrap

One question remaining is how the mapping function is obtained. Similar systems normally assume that it is pre-known or provided by out-of-band mechanisms. In our architecture, using the same rendezvous mechanism, we provide a bootstrap overlay to publish dynamic mappings. Using the mapping for a *well-known key,* a node sends request to a well-known region to obtain the mapping function of a set of services. These mappings however are not expected to change frequently. This introduces more flexibility for providing different mappings for different type of services and changing them when required. (Figure 4)

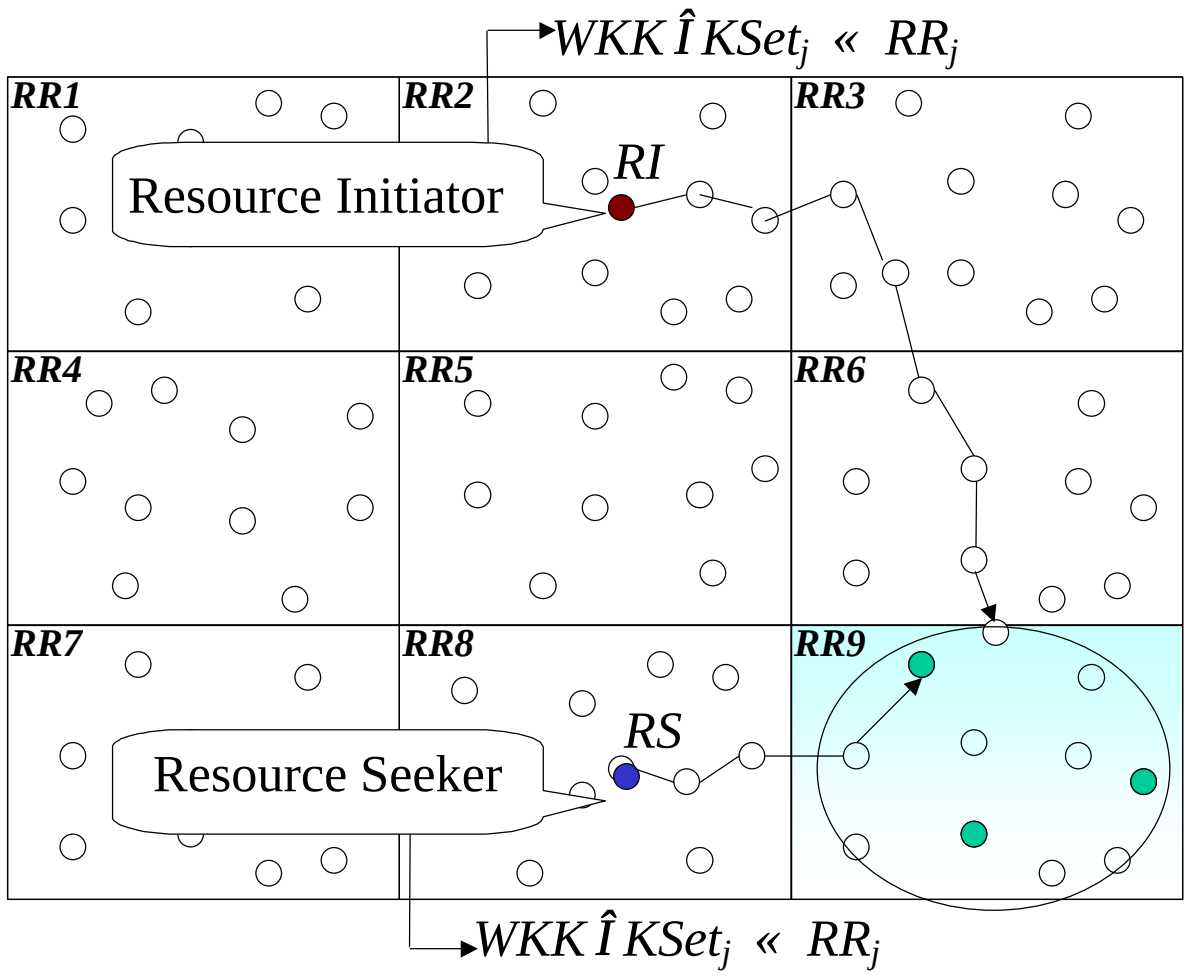

Figure 4: Using the same mechanism for obtaining the mapping function. A bootstrap mechanism for initiating (and obtaining) new resource mappings using *well-known key(s) (WKK).*

## 5 Performance Evaluation

In this section, we evaluate our architecture using detailed NS-2 [15] simulations with detailed models of the wireless MAC and physical layers. We run extensive simulations to investigate the design space, and study the architecture in various environments including node mobility and failures. We study also the effect of inaccurate location. RR is running as an overlay over the routing layer and we are using GPSR as the wireless routing protocol. We modified GPSR to route to regions instead of specific destinations by forwarding the packet toward the center of the region and using geocast or anycast inside the region. We verify the correct operation of RR and evaluate its performance under different scenarios. In addition, we run similar detailed experiments for GHT in order to compare the performance of both systems and identify their characteristics and limitations.

We conducted simulations with up to 400 nodes using the detailed model. To study the scalability of the architecture to higher number of nodes, we perform also higher-level simulations (without the MAC and physical layers) for networks with up to 100,000 nodes and compare it to GHT, flooding (local storage), and centralized storage (external storage).

We evaluate two models, the service model with high lookup-to-insertion ratio (*LIR*) and the event model with low *LIR*. There is a wide range of parameters we consider during the evaluation. The environment parameters include the network dimensions, number of nodes (density), transmission range, and the expected number of queries per second. The design parameters that we study are mainly the number of regions and the number of servers per region. The performance metrics are the success rate of lookups, message overhead per insertion, message overhead per lookup, and total storage/insertion, in addition to the maximum node overhead of these metrics. We also study the overhead due to mobility and failures.

### 5.1 Detailed Simulation Results

We implemented RR in NS-2 as an overlay that runs over any wireless routing protocol that can route towards geographic regions. Currently, we are using GPSR, modified to route to a region, with detailed 802.11 MAC and physical layers. The GPSR beacon interval is 1 sec and the beacon expiration is 4.5 sec. The density is fixed to 1/1024m$^2$ and the number of nodes is 100, 200, or 400. We explored several transmission ranges between 60m and 120m. For convenience (and space limits), in the results shown, we focus on 100-200 nodes topologies and 80m transmission range. The rate of lookups is 2 per second and the number of retransmissions for both insertions and lookups is 3. The periodic failure check interval is 20 seconds. Keys are uniformly distributed at random over the space. The results are the average of 5 random runs over 5 different random topologies. GHT was already implemented in NS-2. The refresh interval for GHT is set to 10 seconds. A common problem in GHT that affects its performance, is when a key hashes to a point outside the external perimeter. In this case, perimeter routing may move around the whole external perimeter during insertions, lookups, or refreshes. To reduce the effect of this problem during evaluation, we modified GHT to avoid mapping the keys to points close to the space boundary. We do that by excluding 10% of each side (left, right, top, bottom) of the space during hashing. We will refer to the modified GHT by GHT*.

#### *5.1.1 Number of Servers in Region*

First, we study the effect of the number of servers in a region. In the simulations, a value is set for the minimum number of servers, below which flooders ask for new servers. Since the server self-election is probabilistic, the actual number of servers in a region could exceed this value. The experiment is run over a 100 node static topology with 4 regions and a LIR equal to 10. The number of insertions varies between 10, 30, and 50. The results obtained are the message overhead/insertion, message overhead/lookup and storage overhead/insertion with different number of servers. From these results we computed the normalized message overhead as *Norm = Ins + Lookup * LIR*, where Ins and Lookup are the overhead/insertion and overhead/lookup respectively. Since, the optimum number of servers depends on LIR (the ratio of lookups to insertions), we show in Figure 5, the normalized overhead for different values of LIR. With low LIR, low number of servers causes lower overhead, while high LIR favors more servers. As we notice, 3-4 servers give a good compromise. Another tradeoff is that the total storage overhead per insertion increases with the number of servers, but also more servers lead to higher robustness in case of mobility or failures.

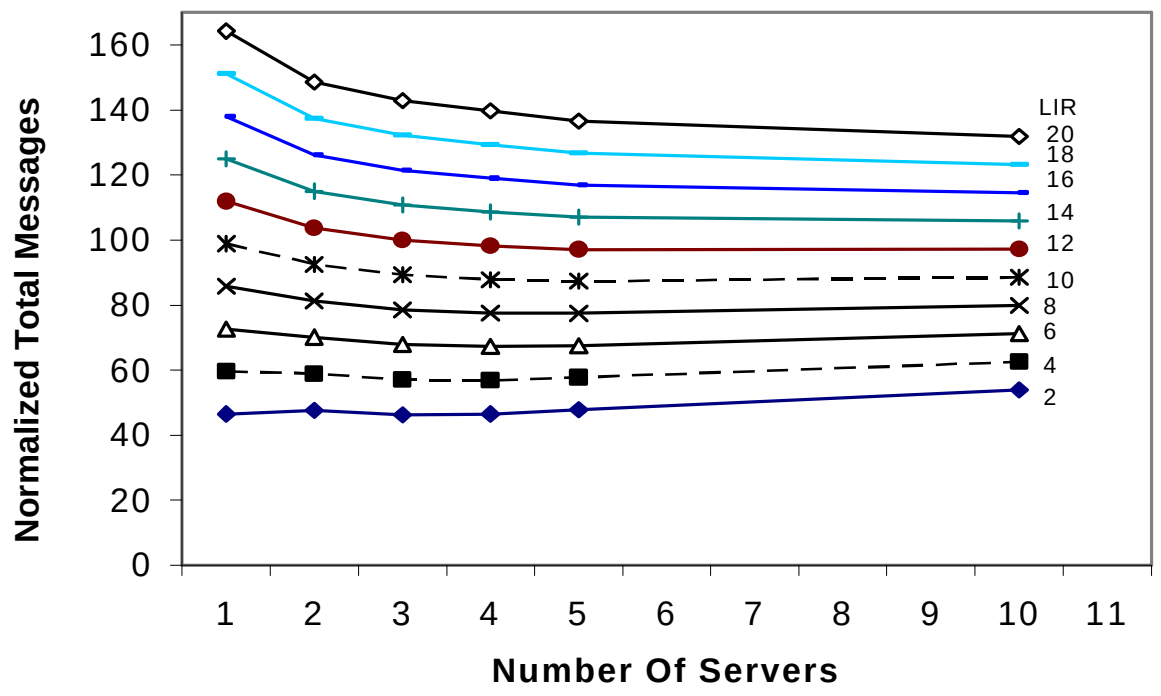

Figure 5: Normalized overhead for different number of servers

*In the coming detailed experiments, the number of servers is set to 3. The simulation run is 200 seconds with 30 insertions at the beginning and 300 lookups at a rate of 2 new lookups per second. Nodes are chosen at random for insertions and lookups. The number of regions varies from 4, 9, 16, to 25. We will start with static networks, then look at failures and mobility.*

#### *5.1.2 Number of Regions*

The number of regions in the space is an important design parameter that indicates the region size and the average number of nodes in region. Ideally we like to have regions with small size to reduce the geocast overhead and at the same time to have the region size large enough to relax the geographic accuracy and reduce the effects of mobility. In static networks, the success rate for both RR and GHT is almost 100%, so we consider only the overheads in this section. Figure 6 shows the insertion overhead of RR compared to GHT and GHT*. RR overhead decreases by increasing the number of regions (i.e. reducing the region size, since the network size is fixed), due to the reduction in the region geocast overhead. At low number of regions, GHT has lower insertion overhead, but they become closer as we increase the number of regions[2]. In Figure 7, the region size has less effect on RR lookup overhead, since the main factor here in reducing the overhead is caching the servers and anycasting them. Both GHT and GHT* have a significantly higher lookup overhead than RR. GHT* has a lower overhead than GHT, since it excludes hashing points close to the network boundary, which can cause the external perimeter traversal, but it is still higher than RR. Insertion and lookup overhead in GHT are similar, since they are using the same mechanism.

[2] Notice that we are using simple direct flooding, where each node inside the region broadcasts the packet once. Alternatively we can use smart flooding techniques [14] which will reduce the geocast overhead significantly.

The high overhead of GHT is due to the home perimeter routing. Each packet in GHT (insertion, lookup, or refreshment) goes to the home node (closest node to destination point), then it traverses the entire perimeter (home perimeter) that encloses the destination, before returning to the home node. The average perimeter length is somewhat long, which can be noticed also from Figure 8, where the average storage overhead for 100 Nodes is around 8 in GHT* (18 in GHT), which means that on average 8 nodes are in the home perimeter and store the packet. This is consistent with the insertion and lookup overheads, which are around 15 messages (including the forwards to home node and back replies). By looking at the randomly generated topology in Figure 9 we can understand why average perimeters are so long: 1) even with avoiding the surrounding boundary points in GHT* and using only the internal 64% (.8*.8) of the space for hashing, there are still many areas where a hash point there could cause external perimeter traversal, 2) even with random uniform distributions there are still somewhat large internal areas without nodes. The keys that hash in these areas have long home perimeters, 3) this may be the most interesting, which is the effect of planarization. Planarization (in this case Gabriel Graph) excludes many links from perimeter routing, such that even in higher dense networks, packets still have to traverse long perimeters because shortcuts between the perimeter nodes are not included in the planar graph[3].

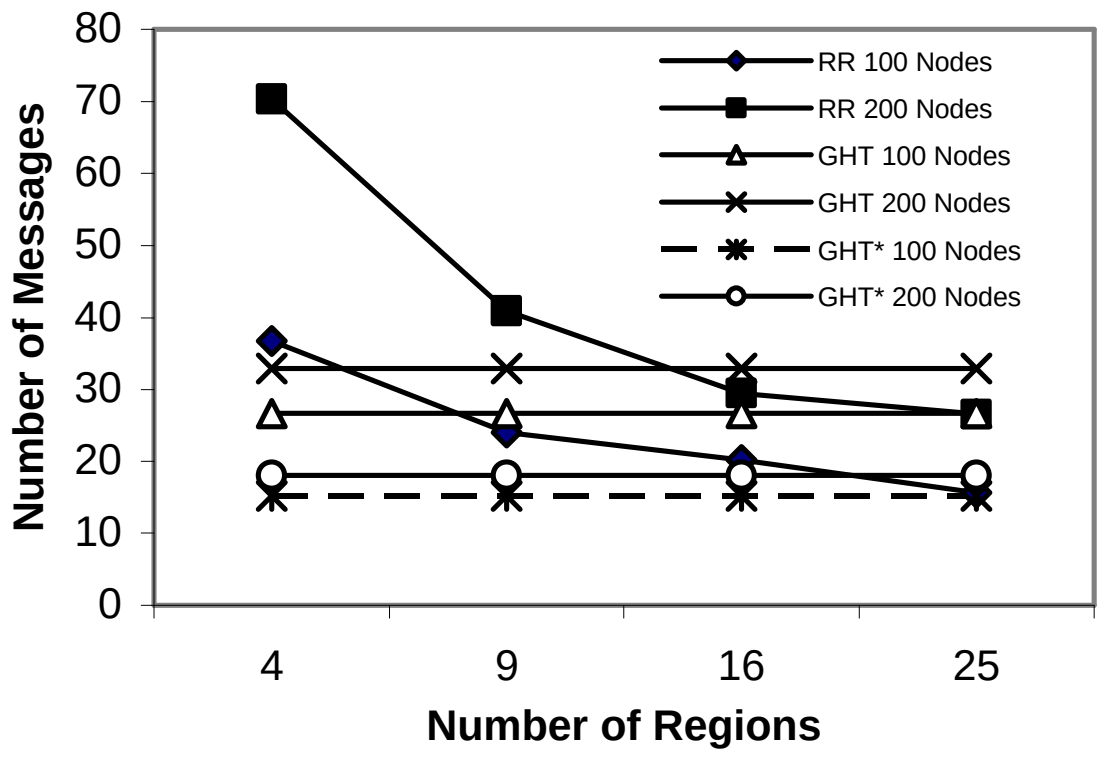

Figure 6: Insertion overhead for RR vs. GHT & GHT*

[3] We are using a density of 1/1024m$^2$ with an 80m transmission range; in GHT [17] the density is 1/256m$^2$ with a 40m transmission range, which gives the same connectivity. In order to verify that, we run the same experiment with the higher density and lower transmission range of [17] and we got the same results.

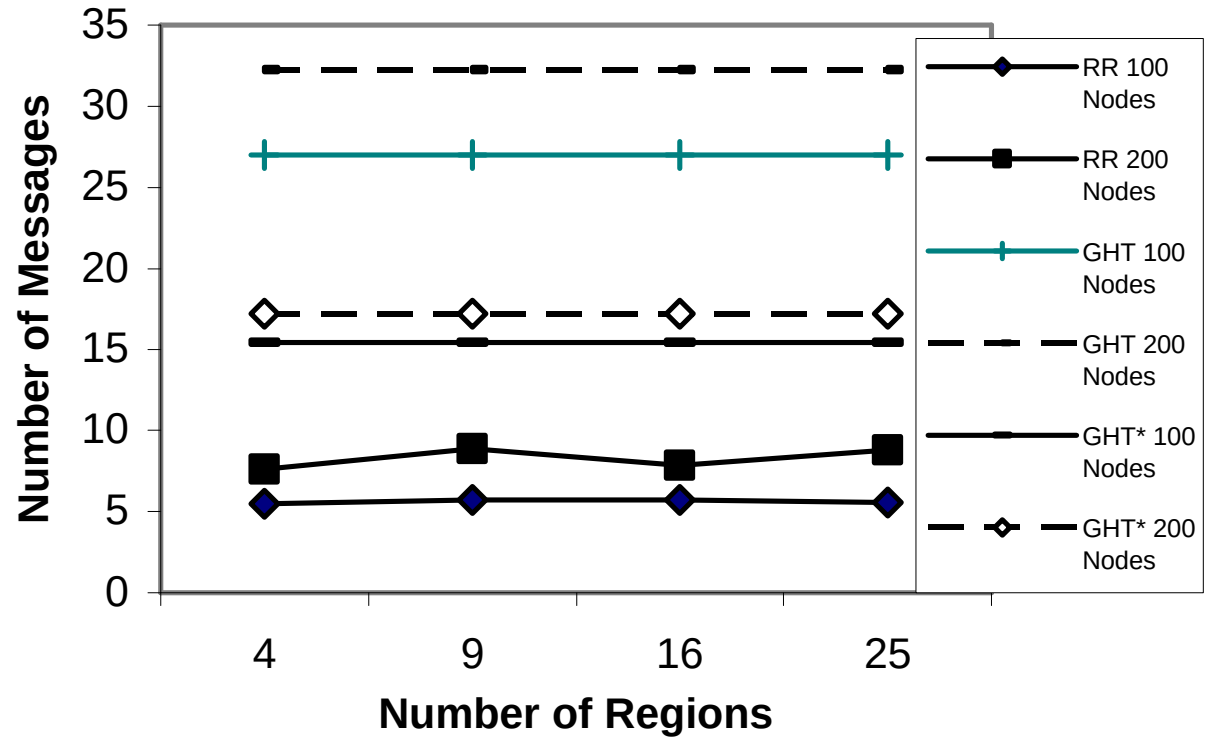

Figure 7: Lookup overhead for RR vs. GHT & GHT*

Figure 10 shows the periodic failure check overhead in RR compared to the refreshment overhead in GHT. In RR it is close to the number of nodes, since in each region, where there are keys stored, one of the servers geocast and so there is only a single geocast in each region independent of the number of keys stored there. In GHT the periodic overhead is much higher, because the refreshment is sent for every key around its home perimeter (every key has a different hash point), so the overhead increases with the number of keys stored. For example, in this case we have 30 keys with an average perimeter of 8 nodes in GHT* (18 in GHT) for 100 node topologies, which gives around 240 messages as shown in the figure (GHT* around 250 and GHT around 500).

These results show that the total overhead depends on the data model and the LIR. For example in the service model, RR has a lower overhead due to the large LIR, while in the event model with larger number of insertions, GHT may have lower overhead. In order to evaluate the total overhead taking into account the insertions, lookup, and periodic overhead, we compute the normalized total overhead as

$$Norm = \frac{Ins}{LIR} * iRate + Lookup * lRate + \frac{Per}{len}$$

where *Norm* is the normalized total overhead per second, *Ins* is the overhead/insertion, *Lookup* is the overhead/lookup, *iRate* and *lRate* are the insertion and lookup rates respectively, *Per* is periodic overhead in an interval, and *len* is the length of the interval. Plotting this equation at different LIRs, we show in Figure 11 the point where RR and GHT* overhead intersect. This is beneficial in deriving which system is more efficient in a certain situation. The overhead ratio between RR and both GHT and GHT* is shown in Figure 12, for different lookup rates (LR). For *LIR>0.1, RR* incurs less overhead than GHT*, with the savings approaching 80% for *LIR>10*.

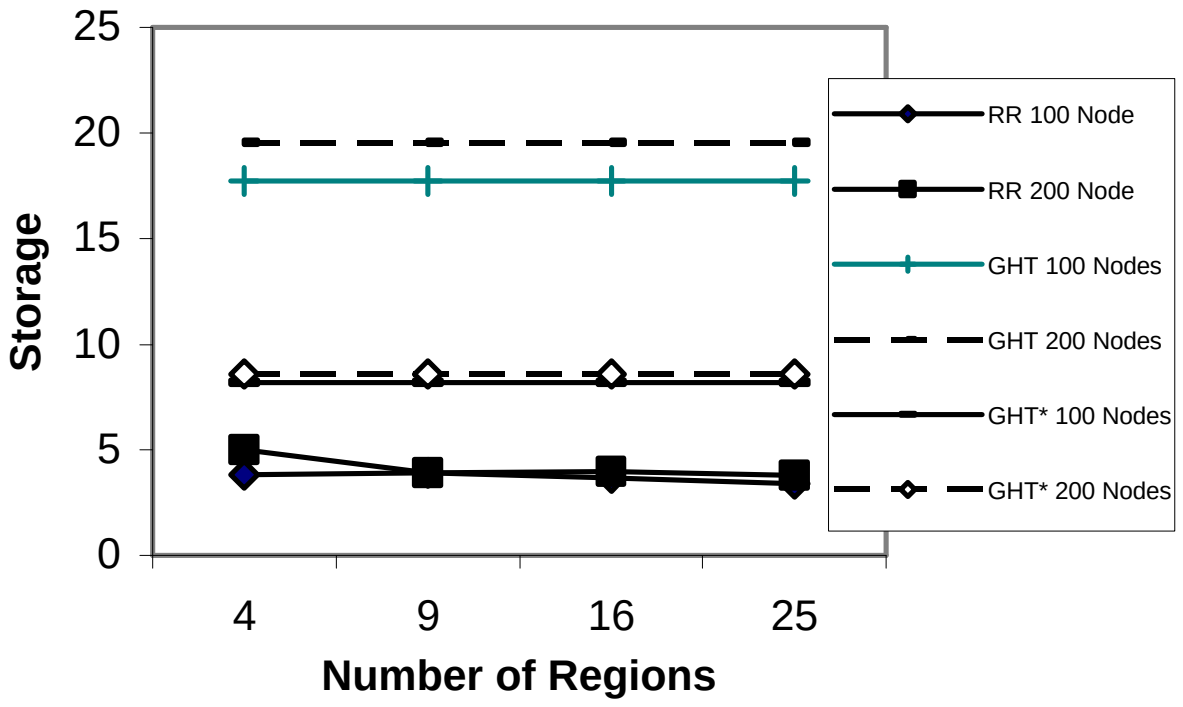

Figure 8: Storage overhead (per insertion) for RR vs. GHT & GHT*

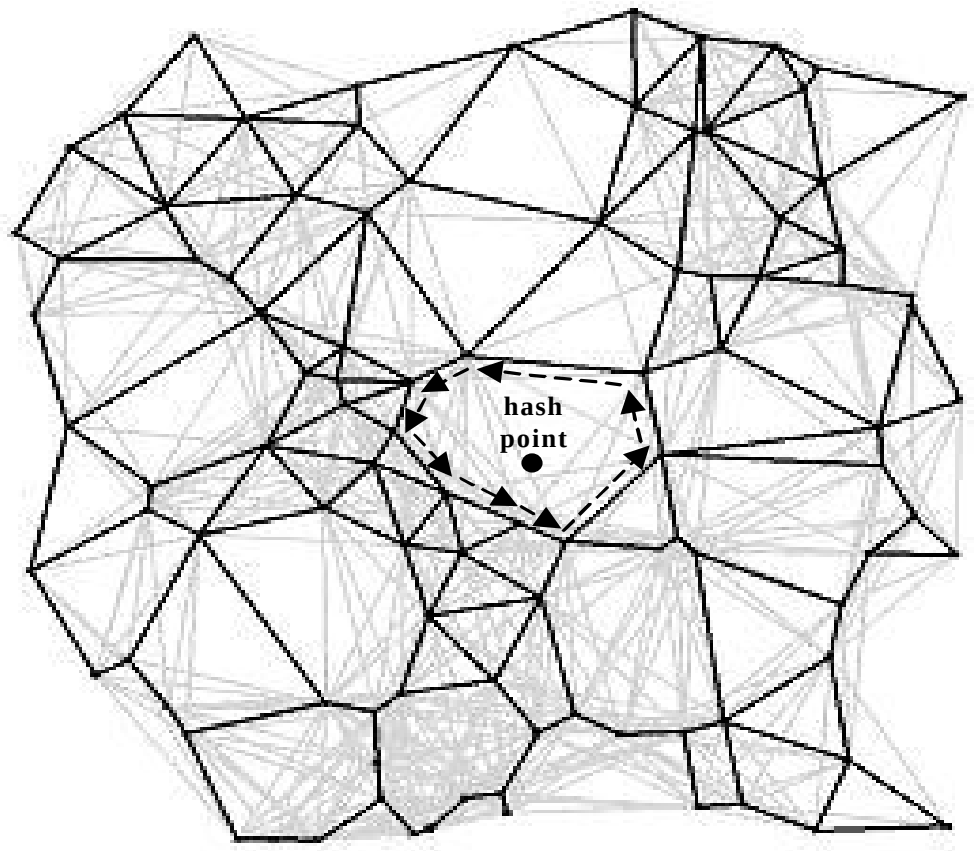

Figure 9: Example of a random uniform topology where perimeter traversing (in a GG planarized graph) takes many hops. Gray edges are the physical links not included in the planar graph

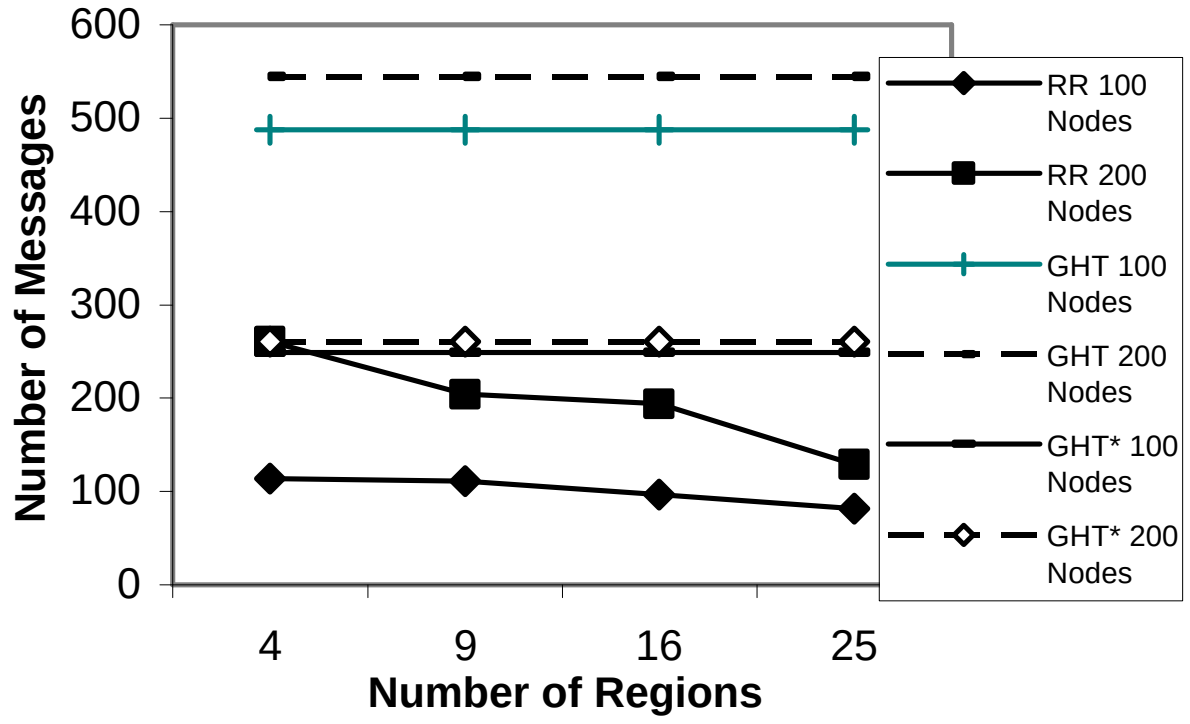

Figure 10: Periodic (refresh, failure check) overhead for RR vs. GHT & GHT*

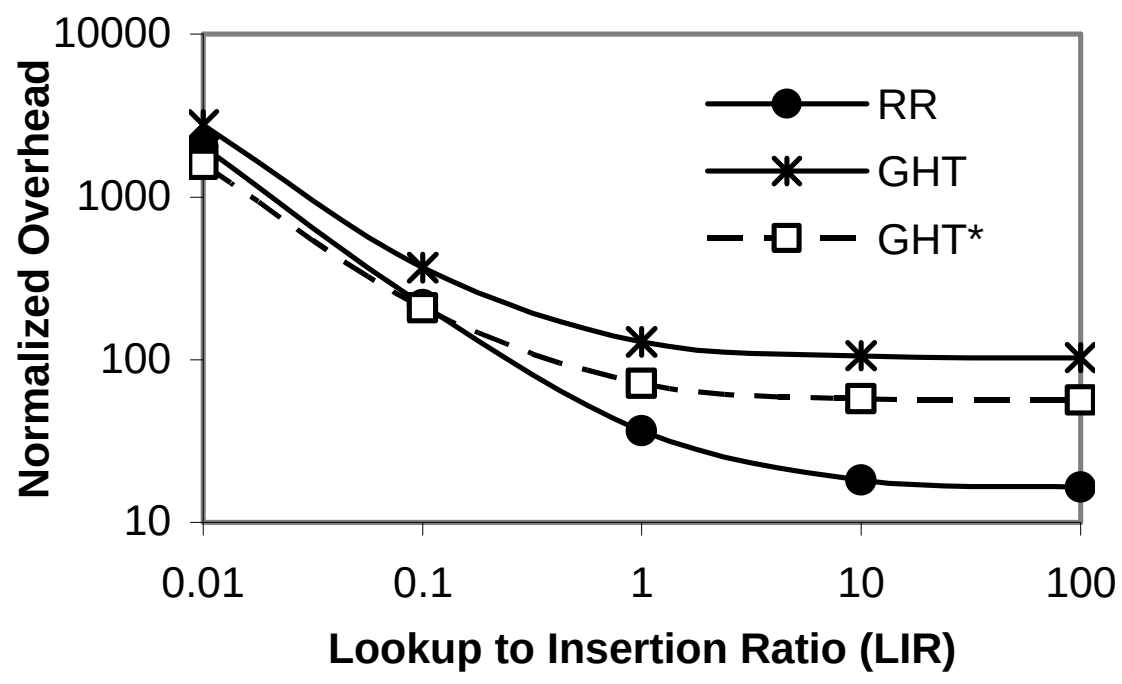

Figure 11: Normalized total overhead per second for different LIR

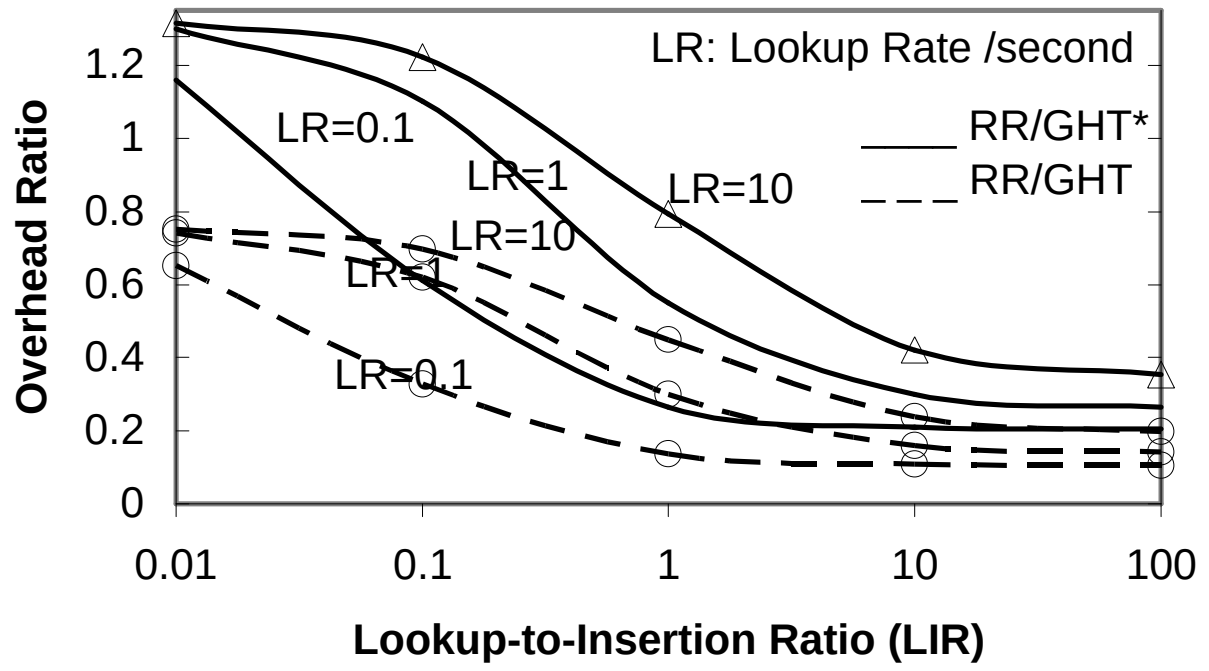

Figure 12: Overhead ratio for different LIR and different lookup rates (LR)

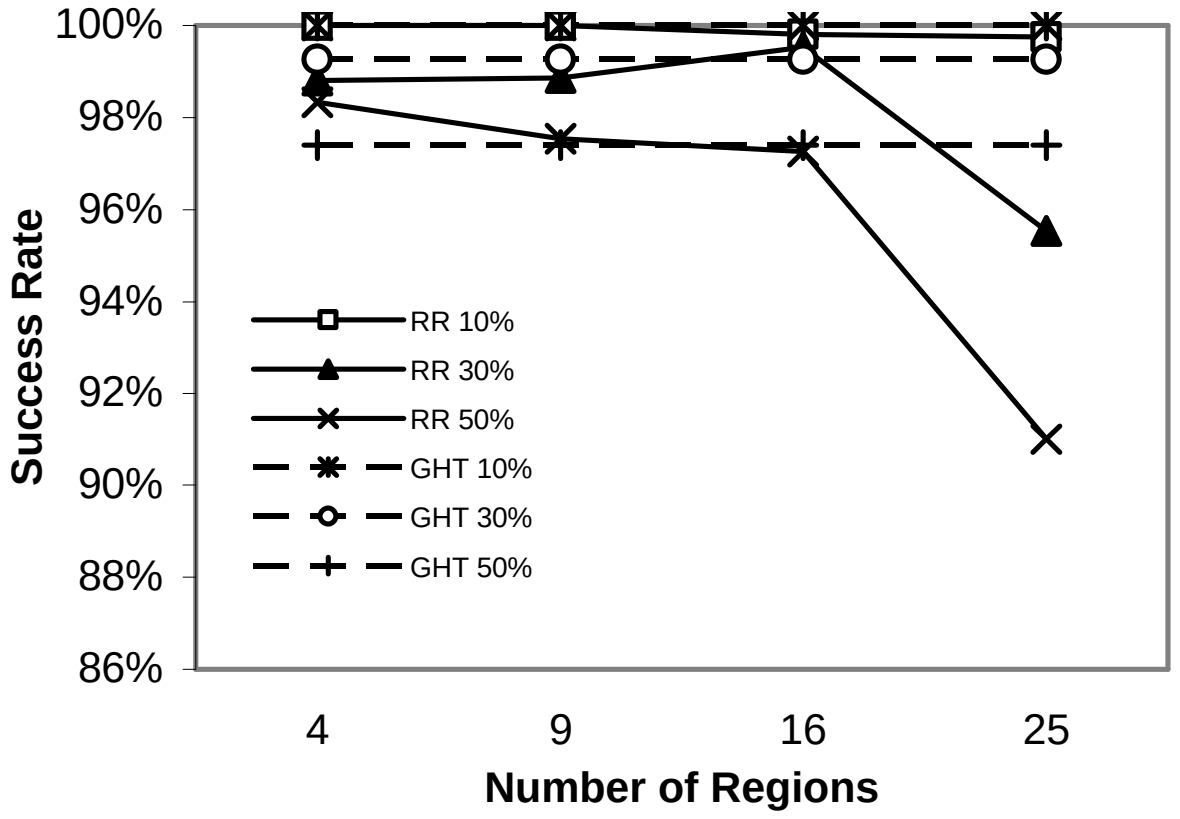

Figure 13: Lookup success rate for different node failure rates

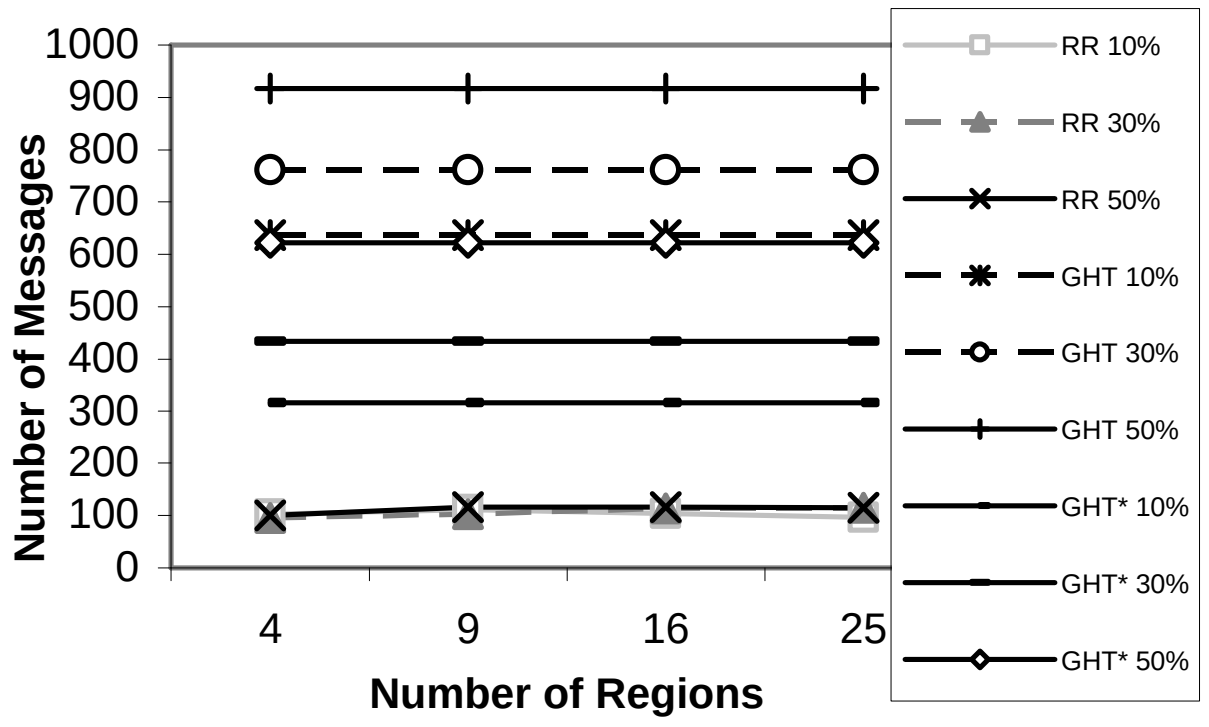

Figure 14: Periodic overhead for different node failure rates

### *5.1.3 Failures*

In this section, we show the robustness of RR with node failures. Robustness is achieved by the replication among servers and the periodic failure checking mechanism. We run simulations of 100 nodes where 10%, 30%, or 50% of the nodes fails at random times. In Figure 13, the success rate remains high with failures in both GHT and RR (around 97% when 50% of the nodes fail). With 25 regions and 100 nodes in RR, the success rate falls down because on average we have only 4 nodes per region, and less than that with failures. By setting the region size appropriately we can avoid this problem. Figure 14 shows the periodic overhead with failures; RR overhead increases slightly but remains around the number of nodes, while GHT overhead increases significantly due to the extra perimeter traversals for key refreshments after node failures.

### *5.1.4 Mobility*

One of the main strengths of RR is its robustness to node mobility. The main reason is that local movements as long as the servers remain in the region, do not require change of servers or any extra overhead. Mobility updates happen only when a server moves out of region. In this experiment, nodes are moving using the random waypoint model [2], with a maximum rate of 1m/s, 2m/s, or 5m/s. There is no pause time, all nodes are moving continuously. Figure 15 shows the high success rate of RR under mobility. It drops only with higher number of regions, because of the low number of nodes per region, which is also the reason for the insertion and lookup overhead increase in Figure 16 and Figure 17, because of the extra geocasts and retransmissions looking for servers in a certain region. This problem does not exist with higher number of nodes and can be avoided by setting an appropriate region size. For example, with 100 nodes moving at 5m/s, 9 regions give a success rate above 99%. GHT success rate drops faster with mobility, since any small movements can cause changes in the key storage. Lookup packets reach home nodes (nodes closest to destination point) that do not have the key due to changes in topology. In addition, perimeter traversal with mobility is susceptible to loops, which may cause the packet to exhaust its TTL. We used a 2 second replanarization timer with GHT to reduce this effect[4]. In Figure 18, we show the mobility update overhead during an interval equivalent to GHT refreshment interval, since both of them reflect the overhead due to mobility (in addition, GHT performs refreshes when new node joins or leaves are discovered). RR has much lower overhead, since only the servers send updates when they move out of region. In Figure 19 and Figure 20, we fix the number of regions of RR to 9 and change the pause time of nodes moving with a maximum random-waypoint velocity of 5m/s. We notice the high success rate and the low overhead compared to GHT and GHT*.

[4] In [17] the TTL was also limited during refreshments to reduce the overhead if looping happens, but we have not included this change here, since it is not clear how the TTL can be set dynamically without extra overhead and what its effect is on other functions of the protocol.

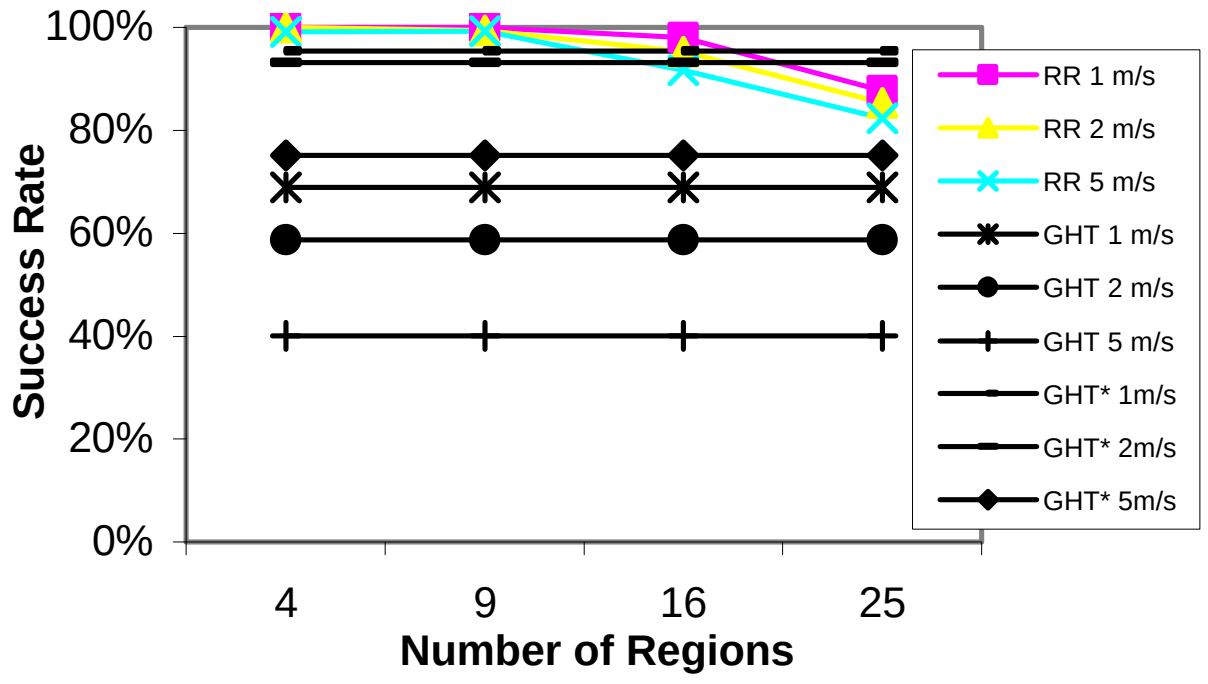

Figure 15: Lookup success rate for different node mobility rates

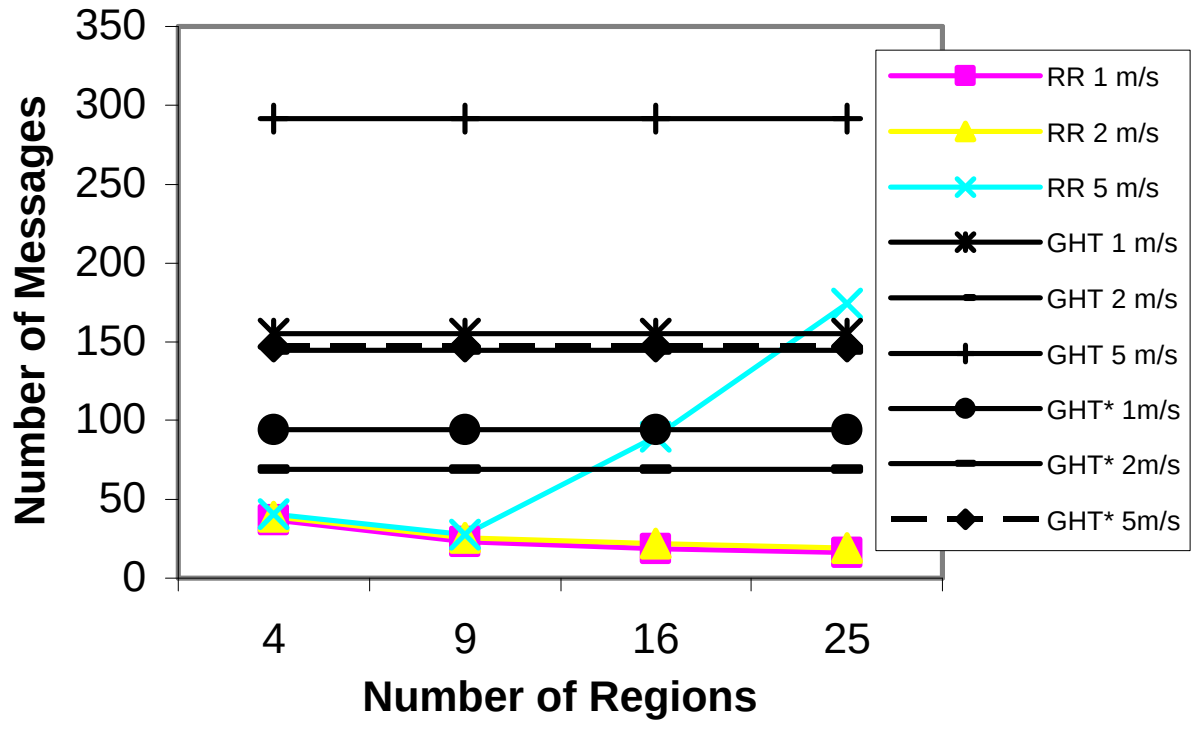

Figure 16: Insertion overhead for different node mobility rates

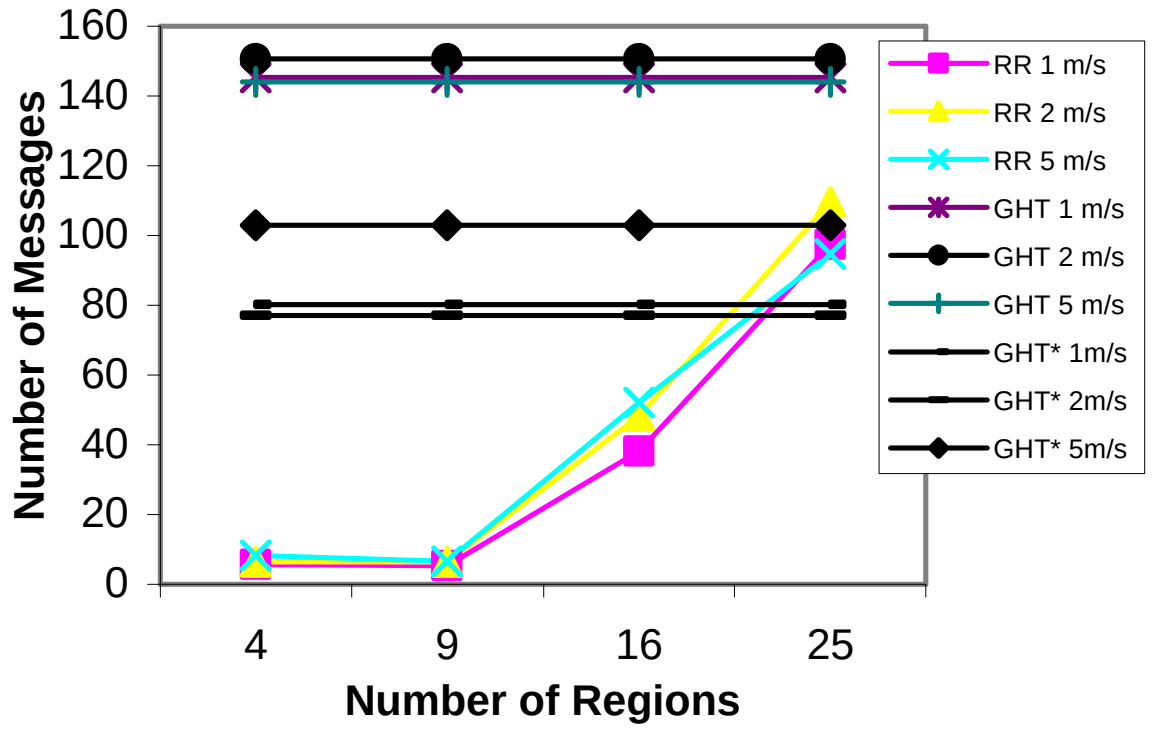

Figure 17: Lookup overhead for different node mobility rates

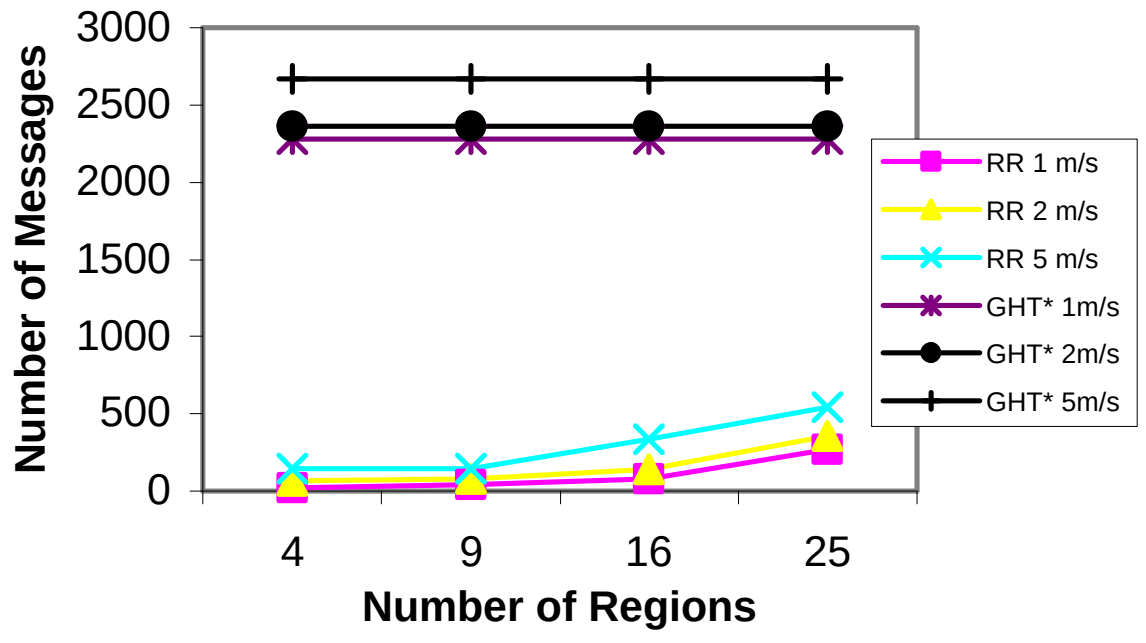

Figure 18: Mobility update (refresh) overhead in RR and GHT

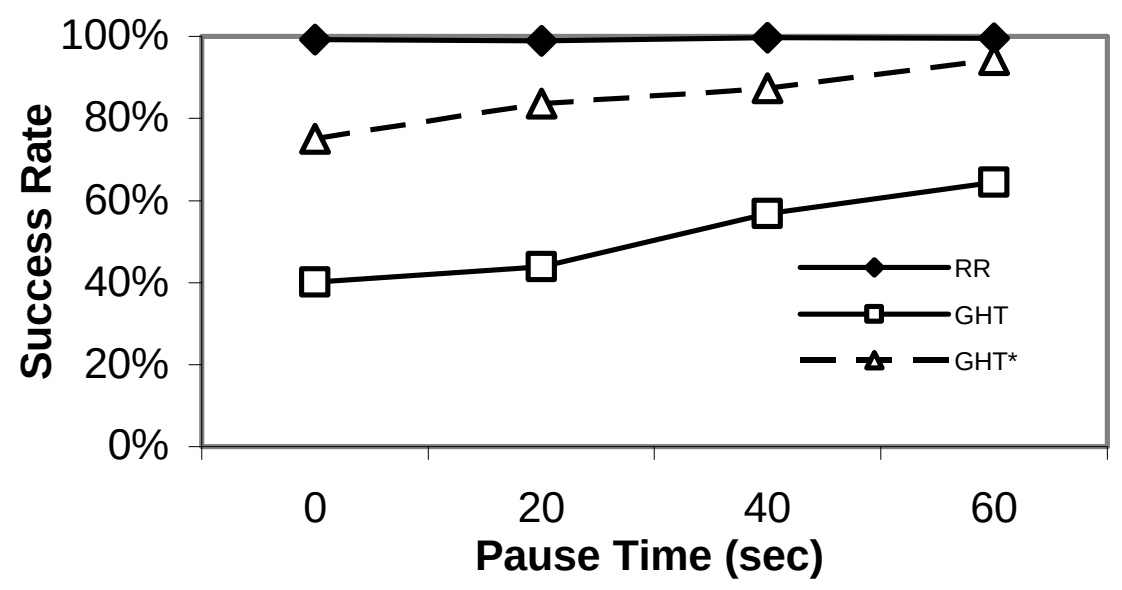

Figure 19: Lookup success rate for different node pause times

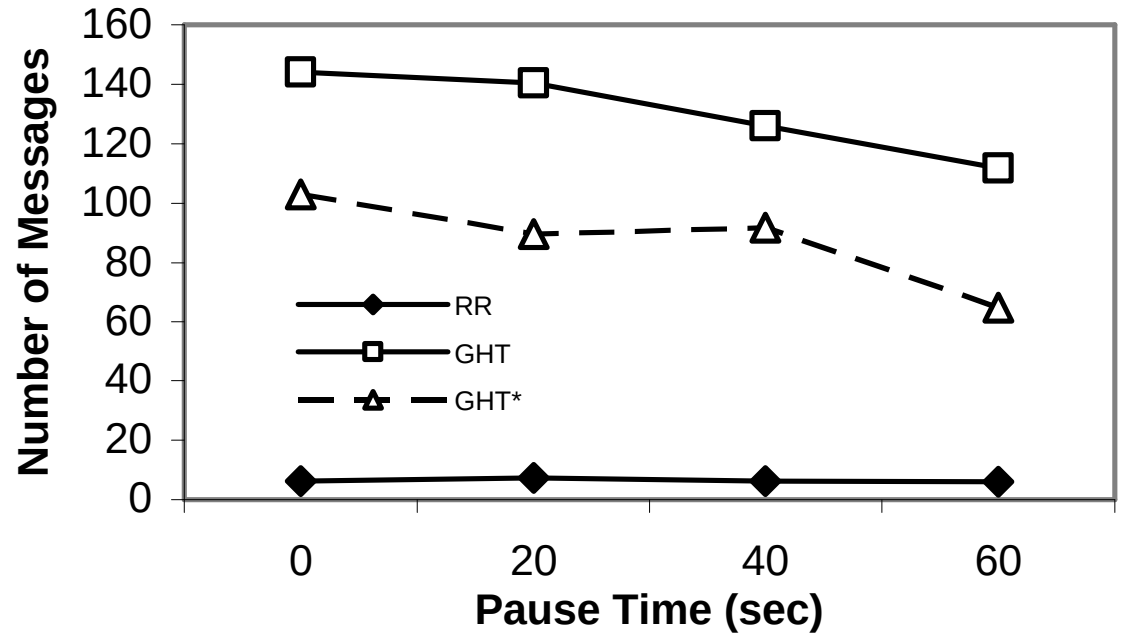

Figure 20: Lookup overhead for different node pause times

## 5.2 Location Inaccuracy

In this section, we study the effect of inaccurate location information on RR. One of the main objectives of RR is to relax the accuracy required by nodes in estimating their location. In this section we consider only the routing behavior in an ideal wireless environment, in order to evaluate the effects of location inaccuracy without the interference from other layers such as MAC collisions or physical layers effects. We use a static and stable network of 1000 nodes having the same radio range and density as in the previous section. Results are computed as the average of 1000 runs, where in each simulation run, nodes are placed at random locations in the topology and 10 insertions and 100 lookups are generated by random nodes. The success rate is the percentage of successful lookups. The maximum localization error is presented as a fraction of the radio range. The estimated node location is picked uniformly from a random location around the node accurate position limited by the maximum localization error.

Figure 21 shows the success rate of RR with different number of regions (16, 36, and 64 regions) compared to GHT at low inaccuracy range (2-10% of the radio range). RR is more robust to location inaccuracy than GHT and the effect of inaccuracy is less on larger regions.

Figure 22 has a higher inaccuracy range (20-100%), which shows the effects of inaccuracy more clearly. For example, at an inaccuracy equal to 60% of the radio range, GHT success rate goes below 60%, while RR is still above 95%.

By adding the fix we introduced in [20] to GHT and the geographic routing in RR, the success rate improves significantly. At low inaccuracy range, both RR and GHT have a success rate higher than 99.8%, with larger regions still having better rate, but the differences become too small. Figure 23 shows the success rate at high inaccuracy range. With an inaccuracy up to 80% of the radio range, both RR and GHT have above 95% success rate. At an inaccuracy equal to the whole range, RR is above 90% and GHT above 80%, which is a significant improvement over Figure 22.

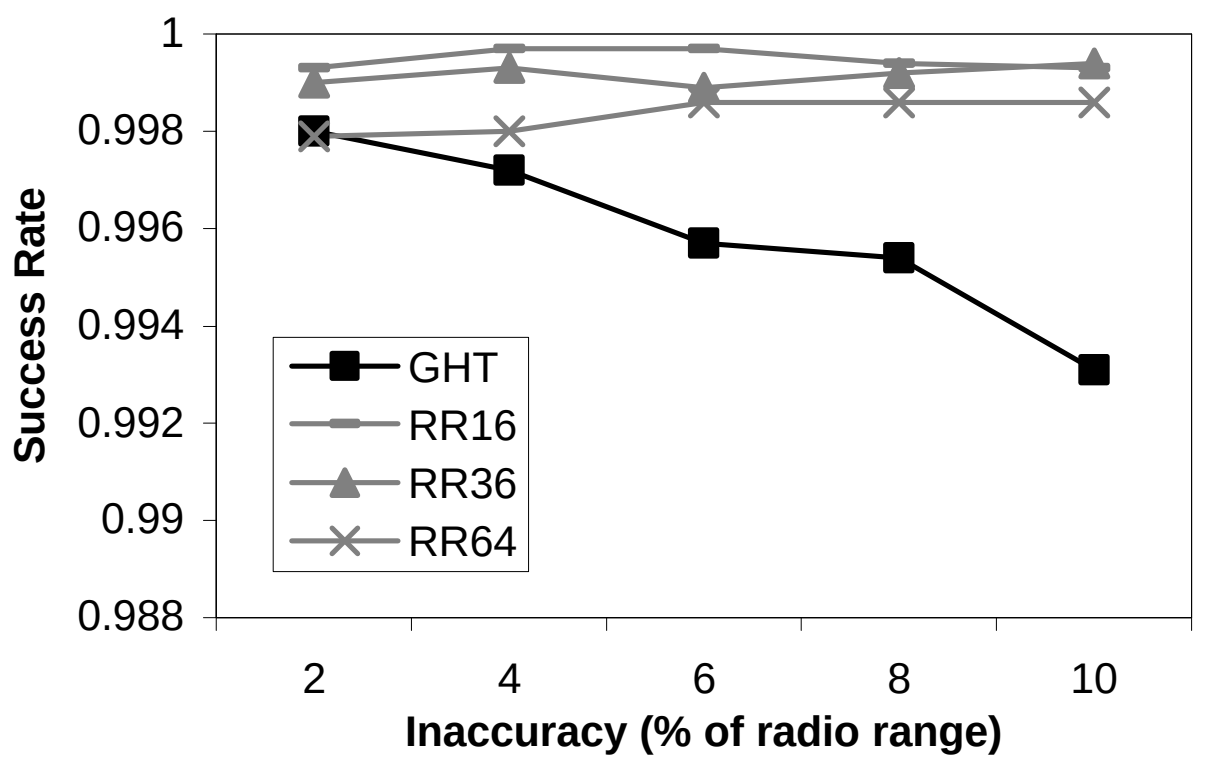

Figure 21: Success rate at low inaccuracy range (2-10% of radio range)

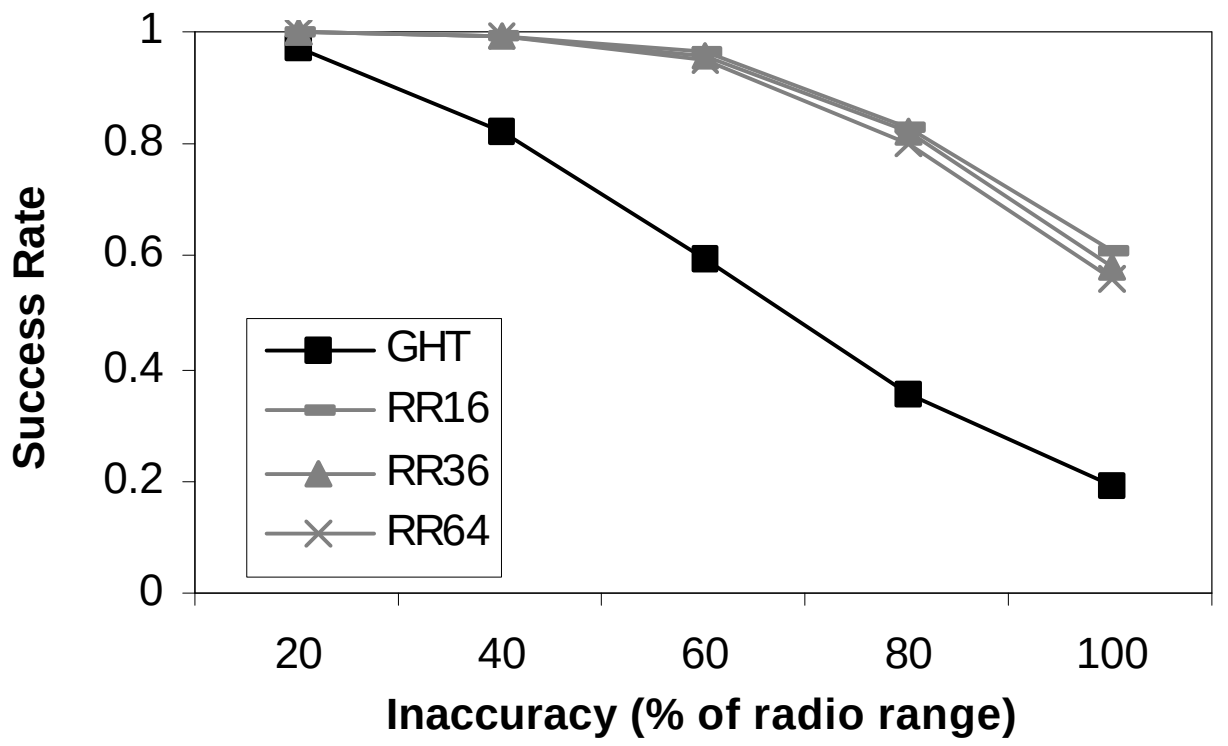

Figure 22: Success rate at high inaccuracy range (20-100% of radio range)

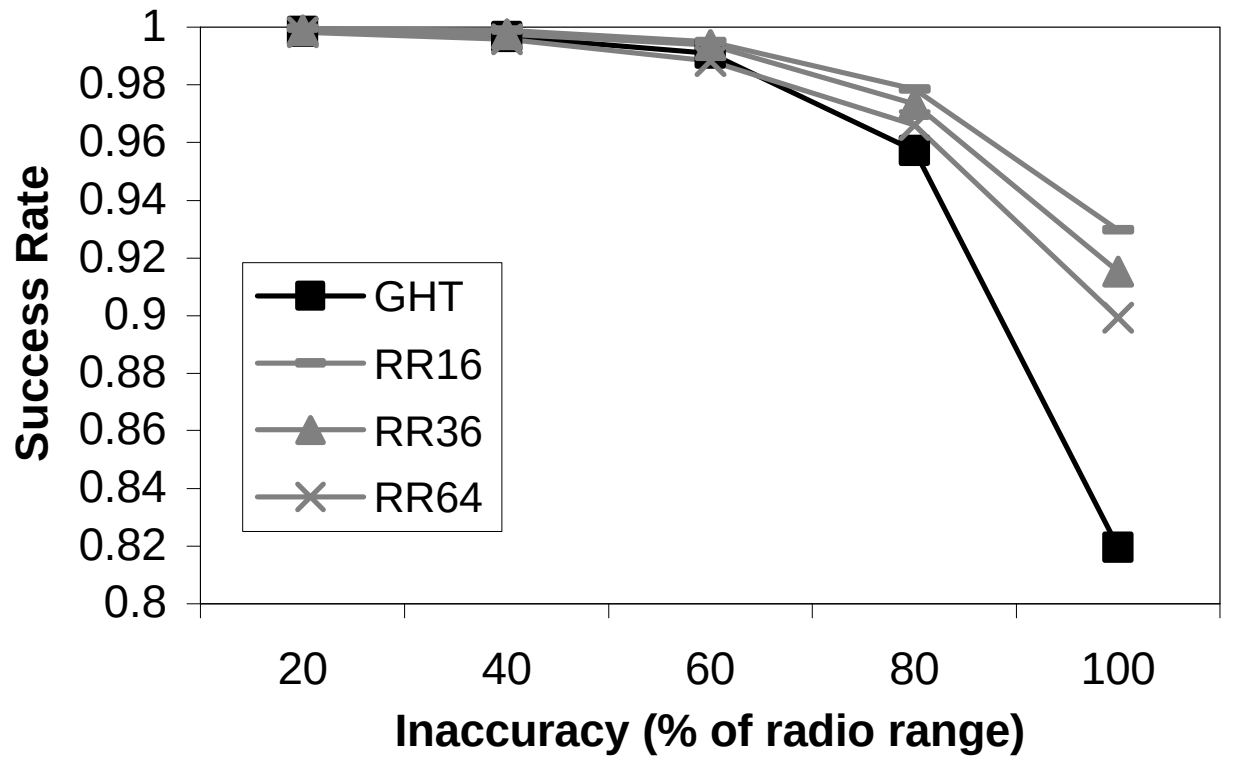

Figure 23: Success rate at high inaccuracy range (20-100% of radio range) with the fix added

### 5.3 High-Level Simulation & Analysis

To evaluate the scalability of RR to higher number of nodes, we perform high-level simulations without the wireless MAC and physical details. We compute the total message overhead and the hotspot message overhead (maximum node overhead) in a static network. In a sensor network, the overhead reflects also the energy consumption of a sensor node and the lifetime of the whole network. Without errors or dynamics, the success rate need not be computed (always 100%). We compare RR to GHT, flooding (local storage), and centralized storage (external storage). We count only the insertion and lookup overhead. The periodic refreshment overhead in GHT and the failure check overhead in RR are not included, but we can easily compute an estimate for them based on the detailed simulations results. In GHT, we do not use hash points that lead to long perimeter traversal (more than square root the number of nodes), in order to avoid the high overhead of the external perimeter traversal.

We will first do some approximate analysis for the communication overhead of the mechanisms. We will consider general insertion and lookup operations, where $n$ is the number of nodes, $I$ is the number of insertions, $L$ is the number of lookups, $R$ is the number of regions in RR, and S is the average number of servers per region. We will use the asymptotic expression of $O(n)$ for flooding the whole network, $O(\sqrt{n})$ for point-to-point routing, and $O(n/R)$ for region geocasts. In flooding we assume that the nodes store (insert) their keys locally and other nodes flood (lookup) to get them. In centralized storage, we assume a centralized node, storing all the keys, so that all insertions and lookups are forwarded to it. The following table shows the asymptotic insertion and lookup message overhead:

| | Total message overhead | Hotspot message overhead |
|---|---|---|
| Flooding | $O(n)\times L$ | $O(L)$ |
| Centralized | $I\times O(\sqrt{n})+L\times O(\sqrt{n})$ | $O(L+I)$ |
| GHT | $I\times O(\sqrt{n})+L\times O(\sqrt{n})$ | $O(\frac{L}{I})$, {for L > I} |
| RR | $I\times O(\sqrt{n}+\frac{n}{R})+L\times O(\sqrt{n})$ | $O(\frac{I}{R}+\frac{L}{\min(I,R)\times S})$ |

The total message overhead in RR includes the insertion overhead, where an insertion is composed of a point-to-point route to reach the region and a geocast inside the region. The lookup is anycast to a cached server, so it can be considered as a point-to-point route. In the hotspot message overhead of GHT, we assume lookups are uniformly distributed over keys so that for each key inserted, its home node will have *O(L/I)* lookup. In RR, we assume also that keys are uniformly distributed over regions and lookups are uniformly distributed over keys, so that the overhead of a server is the insertions of keys in its region and the lookups (anycasted) are distributed between the *S* servers in the region. The term *min(I,R)* takes care of the case when the number of insertions is less than the number of regions, so that lookups are distributed only over those regions that have insertions. In GHT the refresh overhead per interval is equal to *'the number of keys stored (I) * average perimeter length'*. In RR, failure check overhead per interval is at most *'n/R * minimum (I,R)'*, since we do not need to geocast in regions that has no keys. At a constant node density, increasing the number of nodes will increase the network size, and by increasing the number of regions in RR with the network size and keeping the region size fixed, n/R remains constant. In other way, the average number of nodes in the region is constant, since the density and region size are constant. In this case, the asymptotic overhead at large number of nodes for RR will be similar to centralized storage and GHT. We can also see that in the simulations, where we increase the number of nodes from 100 to 100000 with different LIRs. The number of insertions is 10. The density is similar to the detailed simulations and the region size in RR is set to have an average of 100 nodes. The results are the average of 10 random simulations with 10 random topologies. In Figure 24, we see flooding has the highest message overhead since each lookup is flooded. Centralized, GHT, and RR have close total overhead. Figure 25 shows the hotspot message overhead, which is computed as the maximum message overhead at a single node. Centralized and flooding have a high hotspot overhead compared to GHT and RR. RR has lower hotspot overhead than GHT in these scenarios, because the lookups for a key in RR are distributed over multiple servers in the region, while in GHT the same home node get all lookups for a certain key. As we notice also from the table, the hotspot overhead of RR is low when *I* is small compared to *R*. As *I* increases above *R*, the hotspot overhead will increase. For GHT lower LIR is preferred.

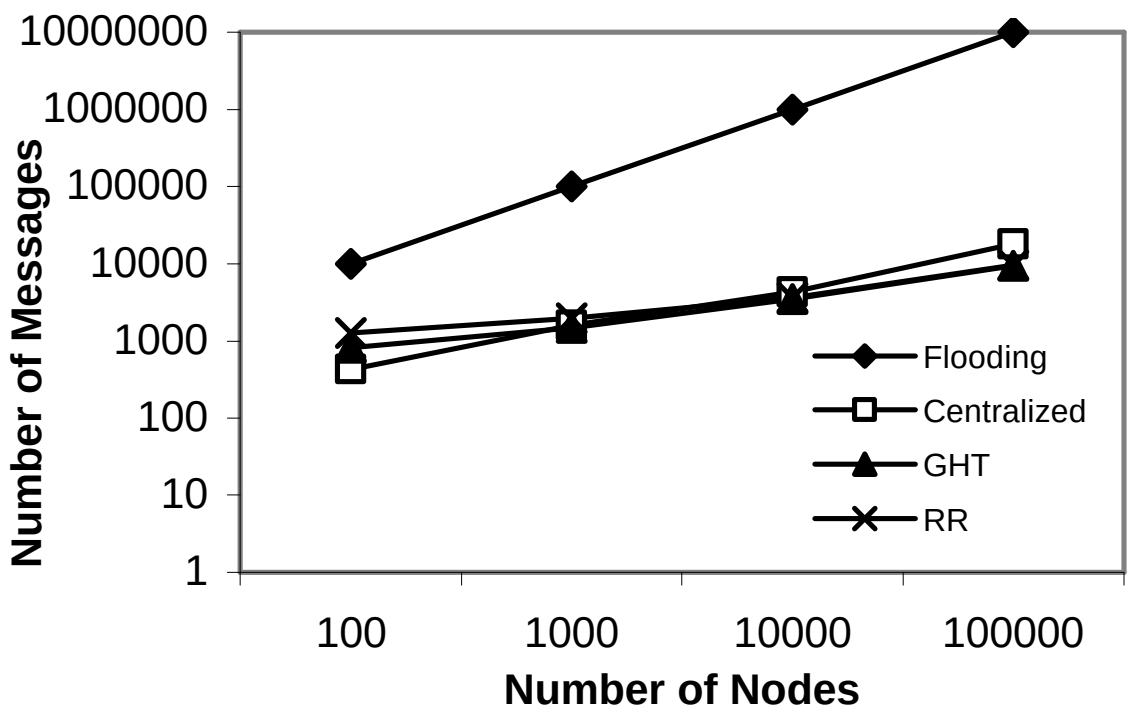

Figure 24: Total message overhead with increasing number of nodes LIR=10

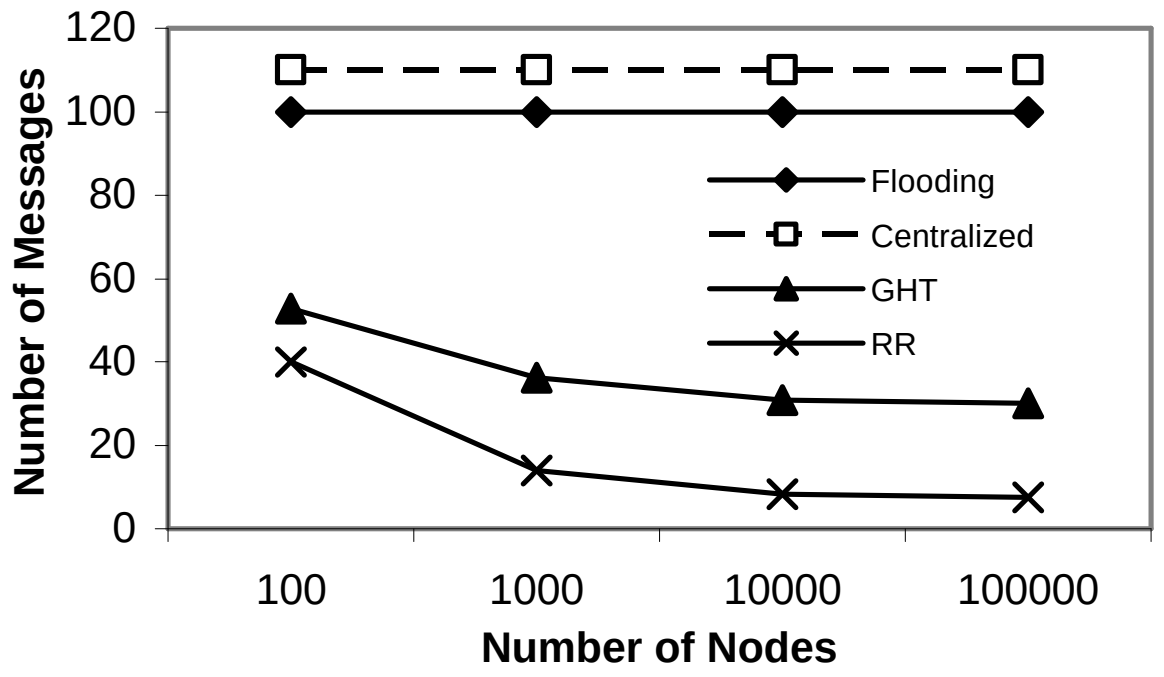

Figure 25: Hotspot message overhead with increasing number of nodes LIR=10

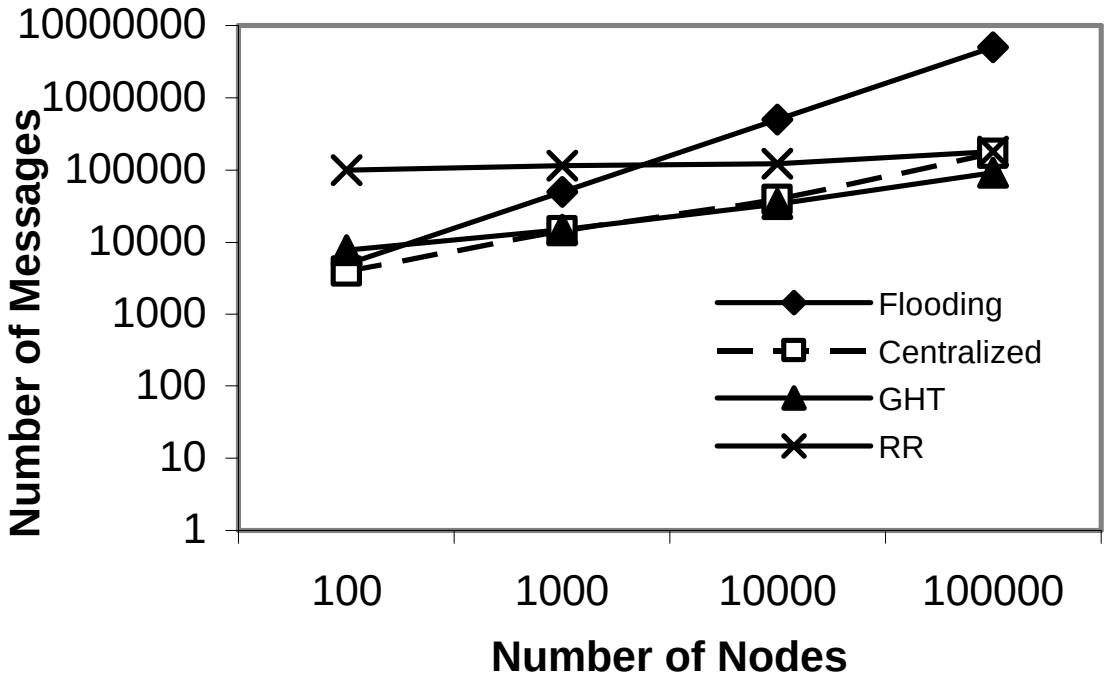

Figure 26: Total message overhead in the event model, LIR=0.05

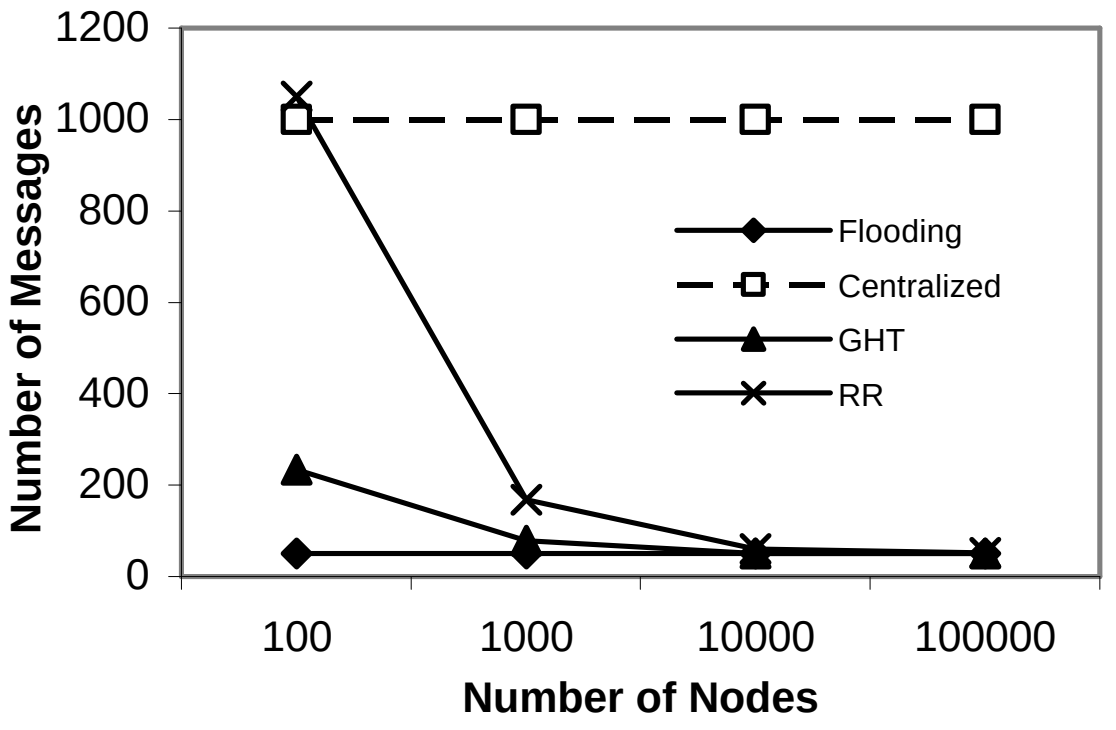

Figure 27: Hotspot message overhead in the event model, LIR=0.05

In Figure 26 and Figure 27, we perform the comparison in the *event model*. We assume 100 event types, 10 detected events per type, and 50 queries. This gives 0.05 LIR. Events are detected at random nodes, while all queries are from a single node representing the network gateway. We assume aggregation is performed at all of them, such that a lookup query returns results as a single reply. We do not consider the structured replication version of GHT. This technique could be used similarly in both GHT and RR. The figures show that RR will have higher overhead at low number of nodes, because of the insertion geocast overhead with large number of insertions. As the number of nodes increase, RR will be similar to Centralized and GHT, since the unicast hops will dominate the overhead in large networks, which confirms the asymptotic overhead shown by the table equation.

# 6 Additional Considerations

In this section, we consider additional design issues: the existence of gaps in the region, empty regions and region resizing.

## 6.1 Region Gaps

As shown in Section 4.3, when a key is inserted in a region, it is stored by all the elected servers in the region. We use geocasting in order to reach all the nodes in the region including the servers. In order to achieve consistency between insertions and lookups, we need a geocasting mechanism that can reach all nodes in the region, otherwise lookups may query servers that are not reached by the insertion geocast. A simple geocasting mechanism is to forward the packet to the region and then flood inside the region boundaries. This mechanism is sufficient at high density networks when all nodes inside the region are connected. But if a gap exists inside the region such that nodes in the region are not connected without going outside the region, this simple mechanism may fail as shown in Figure 28. To solve this problem we proposed a novel geocasting mechanism, Geographic-Forwarding-Perimeter-Geocast (GFPG), that provides guaranteed delivery to all nodes in the region without global flooding or global network information even at low densities and with the existence of region gaps or obstacles. GFPG uses a combination of region flooding and face routing to reach all nodes in the region as shown in Figure 8. For the detailed explanation and evaluation of GFPG and a comparison to other geocasting mechanisms see [19].

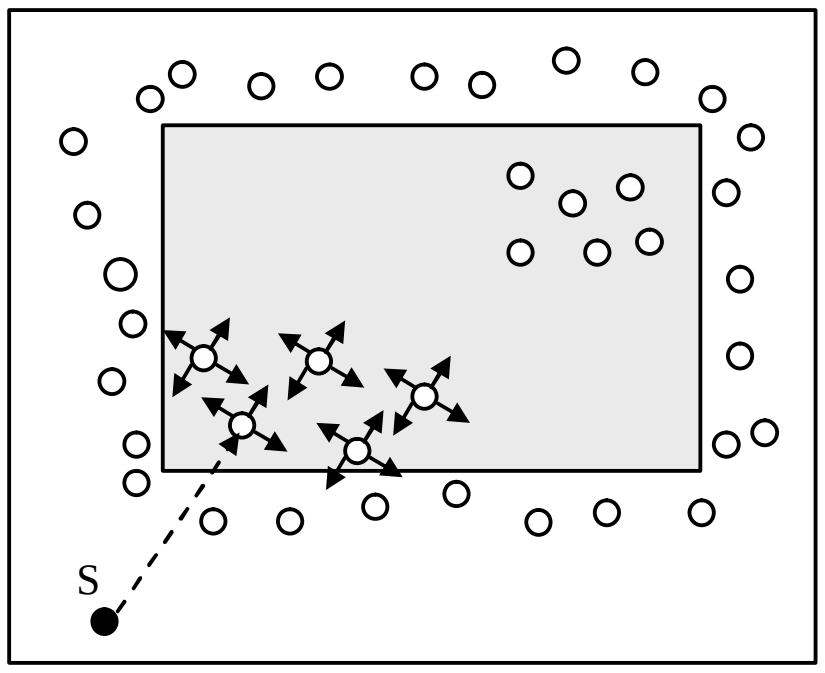

Figure 28: A gap (disconnection) in the geocast region. A packet flooded in the region cannot reach all nodes without going out of the region

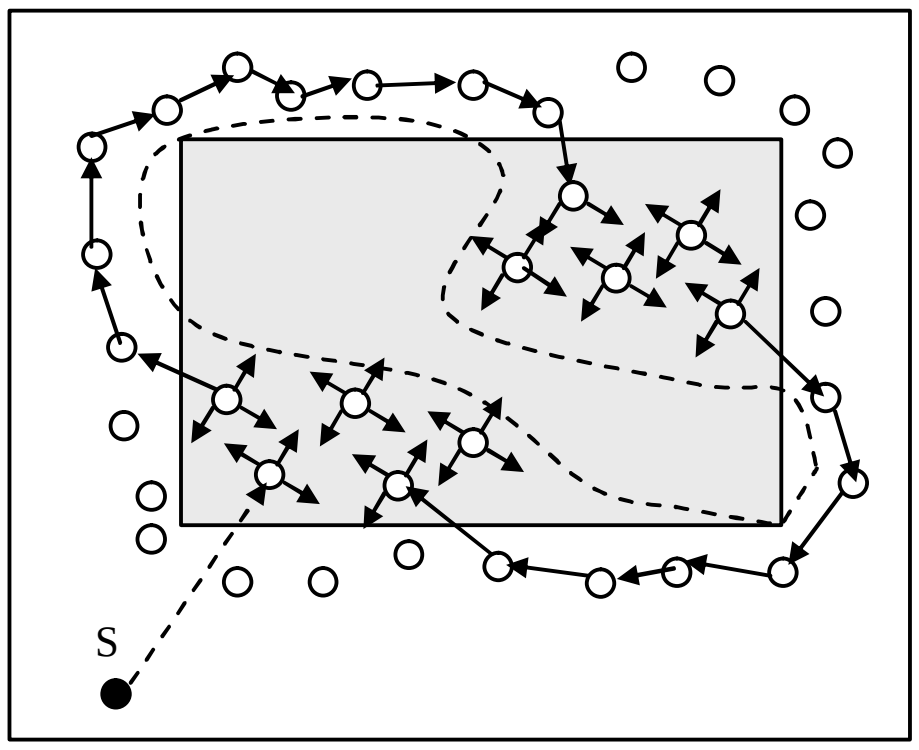

Figure 29: A mix of region flooding and face routing to reach all nodes in the region. Nodes around the gap are part of the same face. For clarity, here we are showing only the perimeter packet sent around the empty face, but notice that all region border nodes will send perimeter packets to their neighbors that are outside of the region

### 6.2 Empty Regions

One aspect of Rendezvous Regions we need to consider is how to deal with empty regions that have no nodes inside. With the relatively large size of regions, this is unlikely to happen when the nodes are uniformly distributed. But for a more general solution that can deal with irregular distributions and obstacles covering a region, we need to consider that. The solution we provide for this problem is to store keys belonging to the empty region in the region of the node closest to that empty region. In both insertions and lookups, a packet forwarded to the empty region will reach the node closest to the region. The node closest to the region will send the packet in perimeter mode, and since there are no nodes closer to the region, it will receive the packet back and it will know that it is the closest node. The node then will insert or lookup the key in its own region similar to regular keys. A server storing keys belonging to other region will have to check periodically that the region is still empty and that no node from other region has become closer to the empty region. If this happens, the key (or a pointer to it) has to move to the other region.

### 6.3 Dynamic Region Resizing

One option we examined is the dynamic resizing of the regions based on the network conditions. We decided not to perform dynamic resizing due to the complexity it will cause and the extra overhead, while the benefits are not clear. We believe the region size should be set as a multiple of the radio range and since the radio range is fixed, then the region size should be fixed. The tradeoff in region size is between the flexibility to mobility and inaccuracy larger regions provide and the low geocast overhead of smaller regions. Typically, a region size covering of a few radio hops is adequate to overcome the effects of inaccuracy and mobility, while our simulations show that reducing the region size too much (closer to the radio range) does not have significant effect on the geocast overhead, since most nodes become border nodes of the region and affect neighbor regions. Accordingly, increasing or reducing the region size beyond a few hops (e.g. 3-5 hops) of the radio range is not beneficial and it makes sense to keep it fixed as a factor of the radio range.

## 7 Conclusions

This paper presents the design and evaluation of RR, a scalable rendezvous-based architecture for wireless networks. RR facilitates service location and bootstrapping in ad hoc networks, in addition to data-centric storage, configuration, and task assignment in sensor networks. We evaluated RR using detailed simulations of a realistic wireless environment including the physical details and node dynamics, and we compared its performance and robustness to GHT. We studied also the scaling properties of RR using high level-simulation and analysis, and compared its scalability to GHT, flooding, and a centralized approach. The results show that RR is scalable to large number of nodes and is highly efficient, especially in applications with high lookup-to-insertion ratios where it provides 80% overhead reduction over GHT. It is also robust to node failures and mobility, and it relaxes the requirements for the geographic accuracy of node positions and network boundaries. In mobile networks, RR has a significant advantage over GHT with higher success rate and much lower overhead, since regions provide a dampening factor to the effects of mobility. Other than the mobility advantages, RR is more robust to location inaccuracy than GHT and it requires lower periodic overhead, since in GHT periodic refreshments need to be sent for each key individually. In addition, RR is more flexible in selecting which nodes to become servers and store the keys. It can choose nodes with certain capabilities from within the region, while in GHT the home node of a key is determined solely by the geography. Selecting more stable nodes with more power and memory can have a significant advantage in sensor networks, where nodes have limited power and resources.